\documentclass[aps,prb,twocolumn,groupedaddress,showpacs]{revtex4}
\usepackage{amssymb}
\usepackage{graphicx}
\usepackage{amsmath,amssymb}
\ifx\pdftexversion\undefined
\usepackage[dvips]{hyperref}
\else
\usepackage{hyperref}
\fi
\hypersetup{
  colorlinks = true, linkcolor = blue
}
\DeclareMathOperator{\sign}{sign}

\begin{document}
\title{Three Mach Zehnder Interferometers    Setup   for Production and Observation of GHZ-entanglement of Electrons
% using a 
}
\author{A. A. Vyshnevyy$^{1}$}
\author{G. B. Lesovik$^{2}$}
\author{T. Jonckheere$^{3}$}
\author{T. Martin$^{3}$}
\affiliation{$^{1}$ Department of general and applied physics, MIPT, Institutskiy per.9, 141700 Dolgoprudnyy, Moscow region, Russia\\
$^{2}$ L. D. Landau Institute for Theoretical Physics RAS, Kosygina str. 2, 119334 Moscow, Russia\\
$^{3}$ Centre de Physique Th\'{e}orique, CNRS UMR 7332, Aix Marseille Universit\'e, Case 907, 13288 Marseille, France}
\date\today
\begin{abstract}
We propose a new single-step scheme for the generation of a GHZ entangled state of three single-electron excitations (flying qubits).
We also present a method to get a generalized GHZ-state. 
Our idea relies upon the most recent progress in the field of on-demand electron sources and mesoscopic Mach-Zehnder interferometry.
We also provide the recipe for the unambiguous detection of this GHZ state via correlations measurements at the output, 
which imply the violation of a Bell-type inequality which is generalized to the case of three particles.
We explain how such measurements can be achieved in the context of Mach-Zehnder interferometry, and draw 
an actual prototype device which could be achieved with point contacts and metallic gates placed 
on a GaAs sample, in the integer quantum Hall effect regime.   
\end{abstract}
\pacs{03.65.Ud, 73.23.-b}
\maketitle
\section{Introduction}
 Quantum entanglement, first noted by Einstein-Podolsky-Rosen(EPR) \cite{EPR} and Schr\"{o}dinger \cite{Schrodinger cat} is a genuine property of quantum mechanics. Qualitative embodiment of this property was 
 given by Bell \cite{Bell original} who showed that entangled states have stronger correlations than allowed by local hidden variable theories (LHVT). Later Clauser and coworkers\cite{CHSH} suggested a more transparent inequality (B-CHSH), the violation of which was experimentally demonstrated for photons\cite{2_photons}.

During the last decade, several works have been achieved in the context of mesoscopic physics to explore two particle entanglement. The initial proposals used superconducting sources of electrons connected to two normal metal leads, through which the two constituent electrons originating from a Cooper pair were split to form a singlet state outside the superconductor\cite{lesovik_martin_blatter,loss}.
Other proposals for two electrons entanglement were subsequently made using ballistic electrons and point contacts placed in the integer quantum Hall effect (IQHE) regime\cite{beenakker_2particle,buttiker_2mz}. Bell inequality tests based on stationary current noise  cross correlations revealed that a maximal violation could be achieved\cite{chtchelkachev,beenakker_2particle,buttiker_2mz,lebedevlesovik,sauret,bayandin}.

In the context of quantum optics, Greenberger, Horne and Zeilinger introduced a maximally entangled tripartite state, commonly called the GHZ-state\cite{GHZ}. The Bell parameter in the tripartite case can take values up to 4, while the corresponding parameter in the bipartite case is no more than $2\sqrt{2}$\cite{Cirel'son}. For both cases the LHVT limit remains 2. In this sense the entanglement is considered to be ``stronger'' than for the two particle state, and it is thus potentially easier for experimental detection.  Ref. \onlinecite{GHZ} pointed out that the measurement results with this state are in conflict with 
local realism when quantum mechanics makes definite (non statistical) predictions, in contrast with EPR states.    
The GHZ state has a number of potential applications in quantum information. It can be used for quantum error correction \cite{NChuang}, and it was proved that tripartite states have advantages compared to bipartite ones in quantum teleportation \cite{tripartite rules 1} and in dense coding \cite{tripartite rules 2}. There are two established methods to produce entangled states of electrons. The  first method exploits the indistinguishability of fermions and relies on postselection to get the desired entangled state\cite{zeilinger_ghz} (see also Ref. \onlinecite{LebedevBlatter} for the  discussion of the role of projection in this case). 
%(this method is used, for example in [ref on Nature ...]), 
The second way is to use the interaction between particles (implication of this method with flying electronic qubits is described in Refs. \onlinecite{koreans,we,another 2 MZ}). 
The latter has the advantage that the evolution during the preparation of the state is unitary and thus can be deterministic with up to 100\% efficiency for producing the desired state. 

While there exist various ways to create and manipulate entangled states of qubits in optical systems\cite{2_photons,3_photons,6_photons} and NMR (nuclear magnetic resonance) 
experiments\cite{shor_nmr}, in mesoscopic systems
the generation of an experimentally determined entangled state of few electrons and
the subsequent proof that Bell inequalities have successfully violated both represent a
considerable challenge. A proposal which follows faithfully the quantum optics 
experimental protocol\cite{zeilinger_ghz} uses edge states in the IQHE and achieves 
the GHZ state using postselection. 

The purpose of the present work is to go beyond this existing protocol and thus to 
explore three particle orbital entanglement with ballistic electrons propagating in mesoscopic devices
using the interaction between electrons. This proposal is motivated by the   
recent progress in achieving single electron sources\cite{glattli_single} and
for buiding effective electron Mach-Zehnder interferometers (MZIs)\cite{mach_zehnder}.
%Our main goal in producing three-electron entanglement is to find out whether the increased strength of correlations can 
%be helpful enough to overcome experimental difficulties concerning the violation of Bell inequalities for 2 particles.

The outline of the paper is as follows. In Sec. \ref{description_section}, we describe the 3 MZI setup and specify
how the beam splitters (BS) and Coulomb interaction operate. In Sec. \ref{GHZ_section} we justify that a GHZ state
is produced at the outcome. In Sec. \ref{Bell_section} we describe the Bell inequality test which is 
used to show that maximal entanglement is achieved. In Sec. \ref{detection_section} we implement this test
for our 3 MZI device. In Sec. \ref{sec:IQHE} we show a possible realization of our setup
in the IQHE regime. We conclude in Sec. \ref{conclusion}

\section{Description of the setup}
\label{description_section}

\begin{figure}[t]
\centerline{\includegraphics[width=7.5cm]{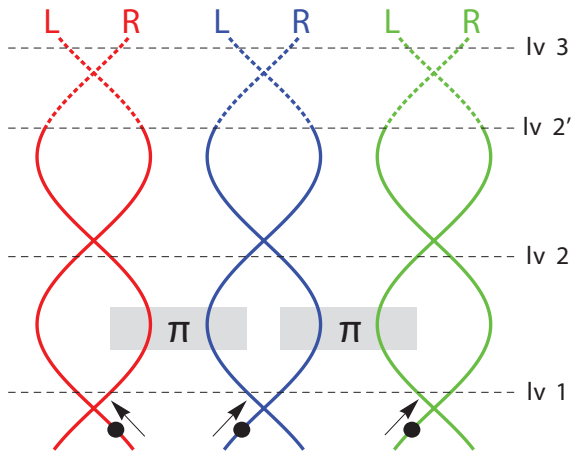}}%setup.eps}}
\caption{(Color online). Proposed setup for the creation of an electronic GHZ-state. 3 MZIs interferometers (full lines), coupled
by Coulomb interactions ($\pi$ boxes). Three electrons are injected simultaneously in the three MZIs,
and propagate from the bottom to the top of the figure.
The crossing of the lines indicate a beam splitter (BS), where an electron can exit
in one of the two output branches. The dotted lines for each MZI show a third BS, used to explain theoretically the
detection process.  The dashed horizontal lines specify different levels of the setup which are 
used in the theoretical explanations.}
\label{setup(f)}
\end{figure}

A schematic drawing of the device is depicted in Fig.~\ref{setup(f)}. 
It consists of three  MZIs, which are placed side by side (shown with full lines in Fig.~\ref{setup(f)}).
Each MZI consists of two incoming channels, which meet at a first beam splitter (BS).
The two outgoing channels  from these BS propagate and meet at the second BS. The theoretical setup also
has a second loop (shown with dotted lines in Fig.~\ref{setup(f)}), with propagation from the output channels
of the second BS, and recombination at a third BS. We will use this second loop to explain theoretically
the detection of the GHZ states, and we will show in Sec.~\ref{detection_section} 
that it is possible in practice to get rid of this second loop and perform the GHZ state production and detection with a single loop. 

We distinguish the right (R) and left (L) side of each interferometer and 
label single electron wave functions accordingly. For instance, at any stage of the 
wave packet evolution, $\psi_{jC}$ denotes an electron wave packet 
on the $C$ side ($C=R,L$) of the $j$-th MZI ($j=1,2,3$).

The most likely candidates for electron channels are
edge states in the IQHE regime, which have the advantage that they are immune to 
backscattering effects by impurities and have long phase breaking time.
Several experiments involving MZI in the IQHE have already been performed 
\cite{mach_zehnder}, some of them involving setups with two MZIs.
A proposal for a detailed setup is given in Sec.~\ref{sec:IQHE}.
We assume that a single electron wavepacket is emitted in each MZI above the Fermi sea. 
Such single electron emission was recently experimentally demonstrated with the on demand single 
electron source\cite{glattli_single,mahe} which uses the mesoscopic capacitor
as the injector.
\begin{figure}
\centerline{\includegraphics[width=4.5cm]{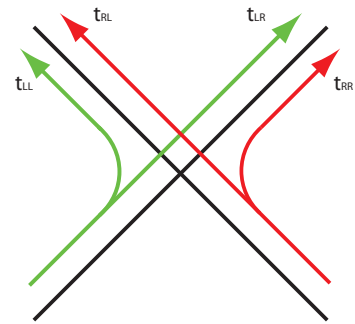}}%setup.eps}}
\caption{(Color online). Reflectionless beam-splitter with transition amplitudes.}\label{fig:BS}
\end{figure} 
The BS are assumed to be reflectionless, i.e. incident particles 
cannot be backscattered in the same channel,
they can only be transmitted further ``up'' (in Fig. \ref{setup(f)}). 
The BS can be parametrized by a transfer matrix 
\begin{equation}
T=\begin{pmatrix}
t_{RR} & t_{LR}\\
t_{RL} & t_{LL}\\
\end{pmatrix}=
\begin{pmatrix}
i\sin{\theta} & \cos{\theta}\\
\cos{\theta} & i\sin{\theta}\\
\end{pmatrix}
\label{eq:Tmatrix}
\end{equation}
which relates incoming states to outgoing states and $\theta$ is the transparency parameter (see Fig.\ref{fig:BS}). For example if $\theta=0$ the beam-splitter is transparent meaning that the incident particle goes from R to L and from L to R without scattering, while for $\theta=\pi/4$ the incident particle may appear in each of output channels with equal probability $1/2$. There exists an additional freedom in choosing the phases of scattering matrix elements but it does not affect results in a crucial manner (see Appendix \ref{BS_appendix} for details). Here we specify the bottom beam-splitters to have 
equal probability $1/2$ for transmission in the $R$ and $L$ side channels ($\theta=\pi/4$). The transfer-matrix is then simply:
\begin{equation}
T= \begin{pmatrix}
i/\sqrt{2} & 1/\sqrt{2}\\
1/\sqrt{2} & i/\sqrt{2}\\
\end{pmatrix}
\end{equation}

This choice of the bottom beam-splitters as well as $\pi$ phase shift in Coulomb interaction is analogous to the studies\cite{koreans, we, another 2 MZ} of two Mach-Zehnder interferometers. In each of them maximally entangled bipartite state was achieved when the first row of beam-splitters was half-reflecting.

The main steps in the production of the GHZ-state are as follows.
The first stage is the synchronized injection of electrons (dots with arrows at the bottom of Fig.~\ref{setup(f)}),
 which then pass through the first BS of each of the 3  MZIs. 
Beyond the first BS, free propagation occurs for electrons wave packets, until 
they reach the ``interaction region'': the $L$ channel of the first MZ and the $R$ channel 
of the second MZ, and separately the $L$ channel of the second MZ and the $R$ channel of the third 
MZI are put in close proximity so that Coulomb interaction effects between the two pairs 
of neighbouring channels become important. The effective 
length of this interaction region is such that it generates an overall $\pi$ phase shift of the 
two particle wave function associated with the two neighbouring channels.  
It may be unclear why Coulomb interaction would produce a phase shift because Coulomb interaction involves energy exchange process. This matter and conditions required for the phase shift to be produced are discussed in Appendix \ref{Coulomb interaction.}.
After the interaction region the three electrons propagate freely towards the second BS, 
but we allow to insert a phase difference $\Phi$ in each MZ loop. The phase differences which are needed here could be achieved via the Aharonov-Bohm effect or via a scalar potential.
Actually Aharonov-Bohm and Coulomb interaction phase
accumulation occur together but logically the
process can be decoupled into sequence of independent processes. From this point of view the Aharonov-Bohm phase effectively changes the transfer matrix of the following BS (see Appendix \ref{BS_appendix}) which is convenient for further calculations.
The final stage of our setup is the measurement of the state of the three particle
in the output arms (top of Fig. \ref{setup(f)}).

\begin{figure}[t]
\centerline{\includegraphics[width=4 cm]{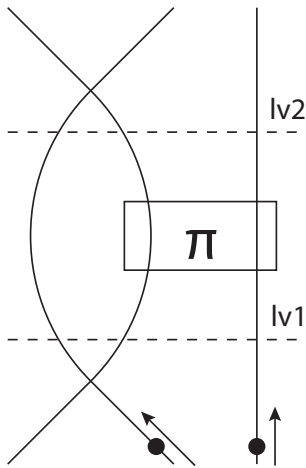}}
\caption{Illustration of the analogy between a MZI and a spin-counter: one MZI coupled to a lead
via Coulomb interaction (see text for details).}\label{one_MZI(f)}
\end{figure}

To understand the role of the interaction regions, 
let us consider first a given MZI with an electron injected into the right lead (see Fig.~\ref{one_MZI(f)}).
 The electron in this interferometer could be treated as a ``flying qubit'', i.e. a two-state system,
  where states are specified by the side chosen in the interferometer, where the electron travels. 
The initial state therefore is $\psi_R$. After the bottom beam-splitter, at the level ``lv1'' (see Fig.~\ref{one_MZI(f)}),
the state of the electron is
\begin{equation}
\Psi = \frac{1}{\sqrt{2}}(i\psi_R+\psi_L).
\end{equation}     
Next, we add a wire coupled to the interferometer with an electron injected simultaneously. So the initial state is $\psi_{1R}\psi_2$. When two electrons pass the interaction zone they accumulate mutual phase $\pi$. After interaction the two-particle state remains separable, and each particle has a wave-function. If the interaction occurs we get at the level ``lv2'' for the left electron
\begin{equation}
\psi_{\pi} = \frac{1}{\sqrt{2}}(-i\psi_{1R}+\psi_{1L}).
\end{equation}
If we do not inject an electron into the right lead then at the level ``lv2'' we have for the left electron 
\begin{equation}
\psi_{0} = \frac{1}{\sqrt{2}}(i\psi_{1R} + \psi_{1L})~.
\end{equation}
%\begin{figure}
%\centering
%\label{pseudospin_counter(fig)}
%\subfloat[Before.]{\includegraphics[width=2 cm]{scheme_2a.eps}}
%\qquad
%\subfloat[After.]{\includegraphics[width=2 cm]{scheme_2b.eps}}
%\caption{Spin flip after electron passing.}
%\end{figure}

\begin{figure}
\centerline{\includegraphics[width=4cm]{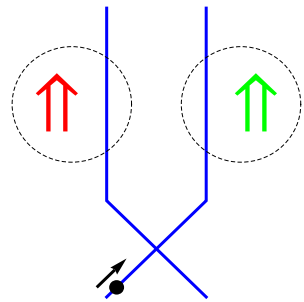}}%scheme_setup.eps}}
\caption{(Color online). The 3 MZI setup (up to lv2) represented in terms of spin-counters: 
because of the interaction regions, the left and right MZIs in Fig.~\ref{setup(f)}
 act like spin-counters (large arrows on the figure) coupled to left and right wire of middle MZI.}
\label{scheme_setup(fig)}
\end{figure}

These two outcomes are orthogonal to each other $\langle\psi_{0}|\psi_{\pi}\rangle = 0$ thus we can distinguish cases when 0 or 1 electrons travel in the neighbouring lead. Moreover the relation $\langle\psi_{0}|\psi_{\pi}\rangle = 0$ is universal in the sense that it is preserved when we change the lead were electrons are injected in the MZI 
or when we place the wire to the left of the MZI 
(as long as the bottom beam-splitter remains half-reflecting). 
We may put into correspondence to the wave function $\psi_0$ a qubit state $|\Uparrow\rangle$ 
and to $\psi_\pi$ a state $|\Downarrow\rangle$.  
This makes our three MZI setup very similar to the situation displayed in Fig.~\ref{scheme_setup(fig)}. There we have a spin-based electron counter with spin flipping after one electron passes through a wire (like it is described in Ref. \onlinecite{Lesuslatter}).
This analogy between MZI and spin-based counter described here is helpful for understanding the nature of the 
GHZ-state in the proposed setup (Fig.~\ref{setup(f)}).  

Up to the level ``lv2'' on Fig.~\ref{setup(f)}, the evolution can be represented in terms of spin-counters (Fig.~\ref{scheme_setup(fig)})
After passing through the beam-splitter the electron state in the middle MZI is $\psi = \frac{1}{\sqrt{2}}(\psi_R+i\psi_L)$. Depending on the arm from which electron came the corresponding spin flips. This produces the final state 
\begin{equation}
\Psi = \frac{1}{\sqrt{2}}(|\Uparrow\rangle \psi_R|\Downarrow\rangle + i|\Downarrow\rangle\psi_L|\Uparrow\rangle),
\end{equation}
which is obviously a GHZ-type state.

\section{GHZ state production}
\label{GHZ_section}
\subsection{Standard GHZ-state production}
We give here the detailed explanation about the GHZ-state production. 
We find an exact expression for the electron state at the level ``lv2'' on Fig.~\ref{setup(f)}, and we identify the local unitary transform making it a standard GHZ state. ``Local'' means that the unitary transform is a direct product of three unitary operators each acting over a Hilbert space of a corresponding electron $U=U_1\otimes U_2\otimes U_3$.
 
Specifically the initial three particle state is chosen to be (see bottom of fig.~\ref{setup(f)}):
\begin{equation}
\psi = \psi_{1R}\psi_{2L}\psi_{3L}.
\label{incoming}
\end{equation}
Note that we choose a simple product state, instead of choosing a Slater determinant of the three 
single particle wave functions. This is due to the fact that the electron wave function do not have mutual parts of trajectories, so that 
exchange effects do not play any considerable role. 

At the level ``lv1'' after passing the first row of beam-splitters:
\begin{equation}
\Psi = \frac{1}{2^{3/2}}(\psi_{1L} + i\psi_{1R})(\psi_{2R} + i\psi_{2L})(\psi_{3R} + i\psi_{3L})
\label{after_first}
\end{equation}
%\textbf{Aharonov-Bohm and Coulomb interaction phase accumulation occur together but logically the process can be decoupled into sequence of independent processes.}
In order to find the state after the Coulomb interaction has acted, we rewrite the previous equation as
\begin{eqnarray}
\Psi &=& \frac{1}{2\sqrt{2}}(\psi_{1L} + i\psi_{1R})\psi_{2R}(\psi_{3R}+i\psi_{3L}) +\nonumber\\ 
     &~&\frac{1}{2\sqrt{2}}(\psi_{1L} + i\psi_{1R})i\psi_{2L}(\psi_{3R}+i\psi_{3L}). 
\end{eqnarray}

The state after interaction (at the lv2):
\begin{eqnarray}
\Psi &=& \frac{1}{2\sqrt{2}}(\psi_{1L} + i\psi_{1R})\psi_{2R}(\psi_{3R}-i\psi_{3L}) +\nonumber\\ 
     &~&\frac{1}{2\sqrt{2}}(\psi_{1L} - i\psi_{1R})i\psi_{2L}(\psi_{3R}+i\psi_{3L}), 
\end{eqnarray}

or in a more symmetrical way
\begin{eqnarray}
\Psi &=& \frac{1}{2\sqrt{2}}(\psi_{1L} + i\psi_{1R})(-i)\psi_{2R}(i\psi_{3R}+\psi_{3L}) +\nonumber\\ 
     &~&\frac{1}{2\sqrt{2}}(i\psi_{1L} + \psi_{1R})\psi_{2L}(\psi_{3R}+i\psi_{3L}), \label{initial state}
\end{eqnarray}
Now it is easy to see the unitary transformation which produces the canonical GHZ-state
\begin{equation}
\Psi_{GHZ} = \frac{1}{\sqrt{2}}(\psi_{1R}\psi_{2R}\psi_{3R} + \psi_{1L}\psi_{2L}\psi_{3L}),
\label{nominal}
\end{equation}
from~(\ref{initial state}). It should transform $\frac{1}{\sqrt{2}}(\psi_{1L} + i\psi_{1R})$ into $\psi_{1R}$, $\frac{i}{\sqrt{2}}(\psi_{1L} - i\psi_{1R})$ into $\psi_{1L}$, $-i\psi_{2R}$ into $\psi_{2R}$, $\psi_{2L}$ into $\psi_{2L}$, $\frac{1}{\sqrt{2}}(\psi_{3R}-i\psi_{3L})$ into $\psi_{3R}$, and $\frac{1}{\sqrt{2}}(\psi_{3R}+i\psi_{3L})$ into $\psi_{3L}$.
This transform is 
\begin{equation}
U=U_1\otimes U_2 \otimes U_3,
\end{equation}
where
\begin{eqnarray}
\label{transformation}
U_1 &=& U_3 = \frac{1}{\sqrt{2}}\begin{pmatrix}
-i & 1\\
1 & -i\\
\end{pmatrix}\nonumber\\
\label{transform}
U_2 &=& \begin{pmatrix}
i & 0\\
0 & 1\\
\end{pmatrix}
\end{eqnarray}

So, at the level ``lv2'' of our setup~(Fig.~\ref{setup(f)}) we have the GHZ-state up to a change of basis achieved by the transform~Eq. \ref{transform}. This transform could be performed by the second BS row between level ``lv2'' and level ``(lv2')''. As one may notice, transformation $U_2$ does not correspond to any $\theta$ parameter in the formula for transfer matrix of beam-splitter (\ref{eq:Tmatrix}). We may avoid this problem by choosing $\theta=\frac{\pi}{2}$ and shifting the AB phase in the next MZ loop by $\frac{\pi}{2}$. In reality we do not need this action because we intend to use ``condensed'' scheme where preparation and measurement are performed with one row of the beam-splitters (see section \ref{section:condensed}).
 
It is interesting to note that from the quantum computation point of view we have an algorithm implemented on 3 flying qubits which is essentually similar to the coding scheme depicted on Fig.10.2. in Nielsen-Chuang\cite{NChuang}.

%%%%% Thibaut : Figures and details related to Nielsen-Chuang have been removed. If we put them we need to say more 
%%%%% about these Figs. and equations to be understandable 
%\begin{figure}
%\centerline{\includegraphics[width=4cm]{Scheme_computation_Nielsen.eps}}
%\caption{Coding scheme}\label{clercor}
%\end{figure}
%
%While it deviates for the latter due to the way that it is practically implemented, it literally has the form depicted in Fig. %\ref{computer.}
%
%\begin{figure}
%\centerline{\includegraphics[width=6cm]{Schrme_computation1.eps}}
%\caption{Scheme for GHZ-state generation.}\label{computer.}
%\end{figure} 

%Quantum gate definitions follow \cite{NChuang}.

%Hadamard gate
%\begin{equation}
%H=\frac{1}{\sqrt{2}}\begin{pmatrix}
%1& 1\\
%1& -1\\
%\end{pmatrix}
%\end{equation}

%S-gate(phase shift)
%\begin{equation}
%S=\begin{pmatrix}
%1 & 0\\
%0 & i\\	
%\end{pmatrix}
%\end{equation}

%and CNOT-gate(Controlled NOT)
%\begin{equation}
%\begin{pmatrix}
%1 & 0 & 0 & 0\\
%0 & 1 & 0 & 0\\
%0 & 0 & 0 & 1\\
%0 & 0 & 1 & 0\\
%\end{pmatrix}
%\end{equation}

\subsection{Generalized GHZ-state production}

In this work we stay focused on the creation of a GHZ-state in its original form $|\Psi\rangle = \frac{1}{\sqrt{2}} (|\uparrow\uparrow\uparrow\rangle+|\downarrow\downarrow\downarrow\rangle)$.
However in some cases it might be useful 
to obtain more general GHZ-type state such as $|\Psi\rangle = a|\uparrow\uparrow\uparrow\rangle+ 
b|\downarrow\downarrow\downarrow\rangle$ with arbitrary $a$ and $b$. Such a generalized state could 
be useful e.g. for quantum error correction.\cite{NChuang} In our setup it is possible to get such a 
GHZ-type state by a slight modification of the above presented method. We simply need to replace the half-reflecting BS in the second MZI with another one. As we pointed earlier in the general case the transfer-matrix of the BS can be represented as follows
\begin{equation}
T=\begin{pmatrix}i\sin{\theta} & \cos{\theta}\\
                  \cos{\theta} & i\sin{\theta}
\end{pmatrix}
\end{equation}
With this beam splitter in the second MZI the electronic wavefunction is: 
\begin{equation}
\psi = \cos{\theta}|\Uparrow\rangle \psi_R|\Downarrow\rangle + i\sin{\theta}|\Downarrow\rangle\psi_L|\Uparrow\rangle),
\end{equation}
which is a generalized GHZ state in a ``rotated" basis. 
Using the same unitary transformation $U=U_1\otimes U_2\otimes U_3$, where $U_i$ are defined in Eq. ~(\ref{transformation}), produces the generalized GHZ-state in its canonical form.

\section{Bell-type inequality for three particles.}
\label{Bell_section}

To detect the GHZ-state experimentally we suggest to use a Bell-type inequality violation test for three particles\cite{UFN_Klyshko}.
 
Let us discuss some basic facts about Bell-type inequalities for 3 particles which constitute one of the possible generalizations 
of Bell inequality on a tripartite case.

We start from an algebraic inequality 
\begin{equation}
|B| = \left|x_1'x_2x_3+x_1x_2'x_3+x_1x_2x_3'-x_1'x_2'x_3'\right|\leq 2,
\label{algebraic}
\end{equation}
which is satisfied when $x_1,x_2,x_3,x_1',x_2',x_3'$ are real variables with absolute values less or equal to $1$
(this is a three particle generalization of the algebraic inequality used for Bell inequality tests on two particles).

Consider a three particle entangled state, written in pseudospin notation 
($R=\uparrow$, $L=\downarrow$, each particle can be detected in one of two leads L and R). The projection of the pseudospin of $j$ on some vector $\mathbf{a_j}$ corresponds to $x_j$ while the projection on $\mathbf{a_j'}$ corresponds to $x_j'$. 
%This procedure is analogous to the one performed in the setup we describe (analogy will be discussed in appendix~\ref{pseudospin analogy}). 
From the point of view of local hidden variable theories, 
after the creation of the three particle state, between the two measurements
the first electron has projections
$\sigma_\mathbf{a_1}$, $\sigma_\mathbf{a_1'}$, the second one $\sigma_\mathbf{a_2}$, $\sigma_\mathbf{a_2'}$, and the third one
$\sigma_\mathbf{a_3}$, $\sigma_\mathbf{a_3'}$ ($\sigma_\mathbf{a_1}=1$ if the spin is parallel to $\mathbf{a_1}$, and 
$\sigma_\mathbf{a_1}=-1$ if it has opposite direction). Different outcomes of measurements are due to a hidden variable $\xi$
which varies from measurement to measurement. We have identified the real numbers $x_1, x_1', x_2, x_2', x_3, x_3'$ with 
$\sigma_\mathbf{a_1}(\xi), \sigma_\mathbf{a_1'}(\xi), \sigma_\mathbf{a_2}(\xi), \sigma_\mathbf{a_2'}(\xi), \sigma_\mathbf{a_3}(\xi), \sigma_\mathbf{a_3'}(\xi)$.
The average value of B is then given by:
\begin{eqnarray}
 \bar{B} &=& \int(\sigma_\mathbf{a_1'}\sigma_\mathbf{a_2}\sigma_\mathbf{a_3} + \sigma_\mathbf{a_1}\sigma_\mathbf{a_2'}\sigma_\mathbf{a_3} \nonumber\\
 &~&~~~~+ \sigma_\mathbf{a_1}\sigma_\mathbf{a_2}\sigma_\mathbf{a_3'} - \sigma_\mathbf{a_1'}\sigma_\mathbf{a_2'}\sigma_\mathbf{a_3'})\rho(\xi)d\xi,
 \end{eqnarray}
where $\rho(\xi)$ is the distribution function of the hidden variable. Experimentally this value can be measured by the following procedure. 
Let us define the correlator
 \begin{equation}
 E(\mathbf{a_1}\mathbf{a_2}\mathbf{a_3}) = \langle\sigma_\mathbf{a_1}\sigma_\mathbf{a_2}\sigma_\mathbf{a_3}\rangle \equiv \int\sigma_\mathbf{a_1}\sigma_\mathbf{a_2}\sigma_\mathbf{a_3}~\rho(\xi)d\xi.
 \label{correlator}
 \end{equation}
Then
\begin{eqnarray}
\bar{B} &=& E(\mathbf{a_1'},\mathbf{a_2},\mathbf{a_3}) + E(\mathbf{a_1},\mathbf{a_2'},\mathbf{a_3}) 
+ E(\mathbf{a_1},\mathbf{a_2},\mathbf{a_3'})\nonumber\\
&~&~~ - E(\mathbf{a_1'},\mathbf{a_2'},\mathbf{a_3'}),
\end{eqnarray}
and from Eq.~(\ref{algebraic}) the local hidden variable average is such that:
\begin{equation}
\bar{B} \leq 2.
\end{equation}
From a quantum mechanical point of view,  the correlator of Eq. (\ref{correlator}) is an average over the state
of an operator describing the measurement of three spins in specific directions:
\begin{equation} 
E(\mathbf{a_1},\mathbf{a_2},\mathbf{a_3})=\langle\hat{\sigma}_\mathbf{a_1}\otimes\hat{\sigma}_\mathbf{a_2}\otimes\hat{\sigma}_\mathbf{a_3} \rangle~,
\label{quant_correlator}
\end{equation}
and
\begin{equation}
\bar{B} = \langle\hat{B}\rangle,
\label{bar}
\end{equation}
where the analog of the ``Bell operator'' for three particles reads
\begin{eqnarray}
\hat{B} &=& \hat{\sigma}_\mathbf{a_1'}\otimes\hat{\sigma}_\mathbf{a_2}\otimes\hat{\sigma}_\mathbf{a_3} + \hat{\sigma}_\mathbf{a_1}\otimes\hat{\sigma}_\mathbf{a_2'}\otimes\hat{\sigma}_\mathbf{a_3}\nonumber\\
&~&~~~~ + \hat{\sigma}_\mathbf{a_1}\otimes\hat{\sigma}_\mathbf{a_2}\otimes\hat{\sigma}_\mathbf{a_3'} - \hat{\sigma}_\mathbf{a_1'}\otimes\hat{\sigma}_\mathbf{a_2'}\otimes\hat{\sigma}_\mathbf{a_3'}.\label{Bell operator def}
\end{eqnarray}

In Appendix \ref{GHZ_appendix} we recall under which conditions this Bell operator
gives a maximal violation of the Bell type inequality. The demonstration is based on the fact that first,
in order for $\hat{B}^2$ to have an eigenvalue equal to $16$, the state must be 
a linear superposition of $| \uparrow \uparrow \uparrow \rangle$  and 
$| \downarrow \downarrow \downarrow \rangle$. Second, the Bell operator
$\hat{B}$ has eigenvalues $\pm 4$ only if the state is related to the
GHZ state:
\begin{equation}
\Psi \equiv  \frac{1}{\sqrt{2}} 
\left( | \uparrow \uparrow \uparrow \rangle + | \downarrow \downarrow \downarrow \rangle \right)
\label{GHZ_state}
\end{equation}
by local unitary transformations. Finally, in this appendix, we specify which angles of spin 
measurement give the maximal value for $\bar{B}$. The resulting angles are:
\begin{equation}
\begin{aligned}
\mathbf{a_1}=\begin{pmatrix}
\cos{\phi_\mathbf{a_1}}\\
\sin{\phi_\mathbf{a_1}}\\
0
\end{pmatrix}&\qquad
\mathbf{a_1'}=\begin{pmatrix}
\cos(\phi_\mathbf{a_1}\pm\pi/2)\\
\sin(\phi_\mathbf{a_1}\pm\pi/2)\\
0
\end{pmatrix}\\
\mathbf{a_2}=\begin{pmatrix}
\cos{\phi_\mathbf{a_2}}\\
\sin{\phi_\mathbf{a_2}}\\
0
\end{pmatrix}&\qquad
\mathbf{a_2'}=\begin{pmatrix}
\cos(\phi_\mathbf{a_2}\pm\pi/2)\\
\sin(\phi_\mathbf{a_2}\pm\pi/2)\\
0
\end{pmatrix}\\
\mathbf{a_3}=\begin{pmatrix}
\cos{\phi_\mathbf{a_3}}\\
\sin{\phi_\mathbf{a_3}}\\
0
\end{pmatrix} &\qquad
\mathbf{a_3'}=\begin{pmatrix}
\cos(\phi_\mathbf{a_3}\pm\pi/2)\\
\sin(\phi_\mathbf{a_3}\pm\pi/2)\\
0
\end{pmatrix},\\
\end{aligned}\label{max viol vecs}
\end{equation}
where $\phi_\mathbf{a_1}+\phi_\mathbf{a_2}+\phi_\mathbf{a_3}=\mp\pi/2$ (different signs identify two classes of angles corresponding to the upper and lower signs in the formulae). The origin of these classes lies in the symmetry of the GHZ-state with respect to reflection in the x-y plane. 
These angles will further be transformed into the corresponding beam-splitter parameters $\theta$ and $\Phi$.

\section{Detection scheme}
\label{detection_section}

\subsection{Detailed detection scheme}
\label{detailed_section}

We are now in a position to describe all the steps for production and detection, 
which are achieved in a rather complicated three MZI setup with each MZI containing double loops
(fig.~\ref{setup(f)}).
%\begin{figure}[t]
%\centerline{\includegraphics[width=7.5cm]{setup2.eps}}
%\caption{Steps for the condensed production and detection of the GHZ pseudospin state}\label{More BS}
%\end{figure}
The incoming state of Eq. (\ref{incoming}) is injected into the first BS and becomes that of Eq. (\ref{after_first}) at level ``lv1''.
The $\pi$ shift is applied to neighbouring channels and it results in the rotated GHZ state of Eq. (\ref{initial state}) 
at level ``lv2''. The true GHZ state of Eq. (\ref{nominal}) is achieved at level ``(lv2')'' by passing through the second beam 
splitter row (with phase shift). 
The next MZI loop is associated with the Bell measurements process. 

To produce the Bell measurement correlator we need three values $x_1,x_2,x_3$ corresponding to separate measurements with results within the band $[-1;1]$. 
In the spin case, the spin projection measurements serve this purpose. 
In our setup we can detect a particle in the left or in the right arm. So one of the possible assignments is $x=-1$ for the particle detected in left arm and $x=1$ for particle detected in right arm. Then the Bell correlator is an average of a product $x_1x_2x_3$ which by definition is

\begin{equation}
E=\langle x_1x_2x_3 \rangle = 1*P_1+(-1)*P_{-1},
\end{equation}
where $P_1$ is the probability for $x_1x_2x_3$ to be equal to $1$ and $P_{-1}$ to -1  correspondingly.
$x_1x_2x_3=1$ in four cases. First one is $x_1=1$, $x_2=1$, $x_3=1$, which corresponds to the case where the three electrons were observed in right arms. 
We define the probability of this event as $P_{RRR}$. Other three possible cases have corresponding probabilities $P_{LLR}$, $P_{LRL}$, $P_{LLR}$. So
\begin{equation}
P_1 = P_{RRR}+P_{LLR}+P_{LRL}+P_{RLL}.
\end{equation}
Analogously
\begin{equation}
P_{-1} = P_{LLL} + P_{RRL} + P_{RLR} +P_{LRR}.
\end{equation}
Finally,
\begin{eqnarray}
E=\langle x_1x_2x_3\rangle &=& P_{RRR} + P_{RLL} + P_{LRL} + P_{LLR}\nonumber\\
~~~&~&- P_{LLL} - P_{LRR} - P_{RLR} - P_{RRL}.\nonumber\\
\label{measproc}
\end{eqnarray}
This correlator measured on level ``lv3'' of the setup of Fig~.\ref{setup(f)} corresponds to the  $\langle\hat{\sigma}_z\otimes\hat{\sigma}_z\otimes\hat{\sigma}_z\rangle$ correlator of the pseudospin.

\textit{Measurement of pseudospin in arbitrary direction.}
As we have seen in Sec. \ref{GHZ_section}, in order to make Bell measurements we should be able to measure pseudospin projections in arbitrary directions, say, $\mathbf{n}$. Nevertheless it cannot be achieved directly because we only can measure the presence of the particles in the right or the left lead (``spin up - spin down"). Instead one can do an equivalent measurement which gives the same expectation values. 
One should transform the state in such a way that the direction $\mathbf{n}$ of the pseudospin converts to $z$ and then perform the standard measure procedure Eq.~(\ref{measproc}). This is done by the upper beam-splitters,  and we obtain: $\langle\hat{\sigma}_\mathbf{n_1}\otimes\hat{\sigma}_\mathbf{n_2}\otimes\hat{\sigma}_\mathbf{n_3}\rangle$
for the state on level ``(lv2')''.
The relation between rotation parameters and beam-splitter characteristics is described in Appendix \ref{BS_appendix}.  
 For each Bell correlator the parameters of this loop are different. As soon as we have defined the correlators $E(\mathbf{a'_1},\mathbf{a_2},\mathbf{a_3}),E(\mathbf{a_1},\mathbf{a'_2},\mathbf{a_3}),E(\mathbf{a_1},\mathbf{a_2},\mathbf{a'_3}),E(\mathbf{a'_1},\mathbf{a'_2},\mathbf{a'_3})$ we are ready to calculate the value of the Bell observable
\begin{eqnarray}
B &=& E(\mathbf{a_1'},\mathbf{a_2},\mathbf{a_3}) + E(\mathbf{a_1},\mathbf{a_2'},\mathbf{a_3}) 
+ E(\mathbf{a_1},\mathbf{a_2},\mathbf{a_3'})\nonumber\\
&~&~~ - E(\mathbf{a_1'},\mathbf{a_2'},\mathbf{a_3'}),
\end{eqnarray}
which is the result of the experiment.

Granted, the fact that we have to include an additional loop in each MZI is quite cumbersome 
and renders the implementation of our setup with mesoscopic devices more difficult. In the next section we will
show that it is possible to perform the same task without the additional loop.

\subsection{Condensed detection scheme}
\label{section:condensed}

Here, we insist that the use of an additional loop in each MZ can be circumvented, provided that
we exploit the  composition of two BS in series. We will call the setup where the additional loop has been removed:
``condensed detection scheme''.
Consider one of three interferometers (Fig.~\ref{setup(f)}). After level ``lv2'' it has two beam-splitters each performing its own unitary transformation. Let the transformation matrices of the BS located between level ``lv2'' and level ``(lv2')'' and the BS between ``(lv2')'' and ``lv3'' be $U$ and $V$ correspondingly. Then total transformation $UV$ is also unitary.
The idea is to perform this total transformation with a single BS, and thus to keep the setup as simple as possible (i.e. 
remove the dotted line part of Fig.~\ref{setup(f)}).

To be precise, note that a single beam-splitter cannot reproduce fully all unitary transformations,
and there is the possibility that the transform $UV$ cannot be reproduced in full generality with 
a single beam-splitter. But it can be shown that, from the  \textbf{point of view of measurement},
it is always possible with a single BS to perform a transformation which gives the same measure as the total transformation 
$UV$. We leave the details of the proof of this statement to the appendices (in particular appendix~\ref{settings_appendix}).

It is interesting to note that in the condensed detection setup,
there exists a measurement scheme where the desired violation $B=4$ is achieved using  
only changes of the phase difference $\Phi_i$, $\quad i=1,2,3$ of the MZI loops,
while the transparencies of the BS remain constant:
\begin{equation}
\left\{
\begin{aligned}
\theta_1 &= \pi/8, \qquad &\theta'_1 &=& \pi/8,\\
\Phi_1 &= -\pi/2,   \qquad &\Phi'_1 &=& \pi/2,\\ 
\theta_2 &= \pi/4, \qquad &\theta'_2 &=& \pi/4, \\
\Phi_2 &= \pi,     \qquad &\Phi'_2 &=& \pi/2, \\
\theta_3 &= \pi/8, \qquad &\theta'_3 &=& \pi/8,\\
\Phi_3 &= -\pi/2,   \qquad &\Phi'_3 &=& \pi/2,\\
\end{aligned}
\right.
\end{equation}
Here we stress that the $\Phi$ phases presented in the last formula are ``effective",
 meaning that they depend on the practical implementation of the beam-splitters 
 (values in the formula are for $T$ given by Eq.~(\ref{eq:Tmatrix})). 
 In Appendix~\ref{BS_appendix} we show how to adjust the phases in the case of real experiments. 
This constitutes the justification for removing the additional loop in each MZ. The whole set 
of measurement schemes with $B=4$ is presented in Appendix~\ref{settings_appendix}.

\section{Setup in the IQHE}
\label{sec:IQHE}
While the setups which we have presented and discussed in this work were schematic,
we think that a practical implementation is in reach with current experimental techniques.
Using edge states in the integer quantum Hall effect regime (IQHE), one can obtain chiral channels
where backscattering is impossible. Beam splitters can be obtained using quantum point contact (QPC).
Several experiments realizing one or two electronic MZI have already been achieved~\cite{mach_zehnder}.
 
\begin{figure}
\centerline{\includegraphics[width=7.5cm]{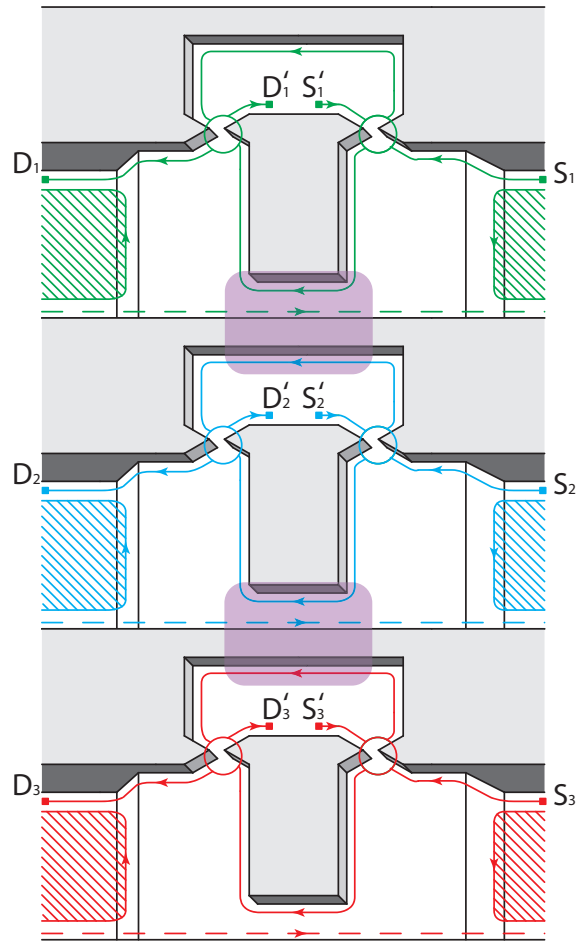}}
\caption{(Color online). Geometry of the setup in the IQHE regime. 
 $S_j, S_j'$ ($j=1,2,3$) are electron sources and $D_j,D_j'$ are drains. Guiding centers of electron edge states meet
 at quantum point contacts, whose transparency can be controlled with gate voltages. The rounded rectangles indicate
 the interaction regions between neighbouring MZIs. Space available for the Fermi sea is within the hatched regions}\label{practical}
\end{figure}

 The actual geometry for a working device, reproducing the setup of Fig.~\ref{setup(f)} is shown in Fig.~\ref{practical}.
This picture represents a 3D potential landscape. Lines going near the potential walls represent chiral edge eigenstates in IQHE. Dotted lines represent unoccupied edge states. The Fermi sea is expulsed from the setup region to prevent screening effects and parasitic entanglement. 
The edge states meet at QPC's, whose transparency can be controlled by gate voltages.
Each MZI has two electronic sources (noted $S_i$ and $S_i'$, $i=1,2,3$, on Fig.~\ref{practical}),
 and two drains collecting the electronic current($D_i$ and $D_i'$). Note that by changing gate voltages,
 it is possible to modulate the length of the different paths, and thus to control the Aharonov-Bohm phase acquired
 by the electrons during propagation. The two circled rectangles on the pictures show the interaction regions, where two branches of two 
 different MZIs are put close to each other, in such a way that a $\pi$ phase shift is produced when electrons are present in the two branches.
 
In order to have time-correlated propagation of electrons over interaction regions, we need on-demand electron sources similar to the ones which have already been experimentally realized in the IQHE \cite{glattli_single,mahe}. 
 In experiments on single-electron injection time of injection varied from $10^{-10}\,s$ to $10^{-8}\,s$ (in this paragraph we use numerical values from Ref. \onlinecite{glattli_single,mach_zehnder,mahe}). The wave packets should be shorter than the length of  interaction region. The mall size of wave-packets ensures that the interaction produces a simple phase shift of the whole wave packet (see Appendix \ref{Coulomb interaction.}). The suitable interaction time $t$ then should be at least 10 times larger (more detailed calculations in the case of two MZI are presented in Ref. \onlinecite{we}) and starts with $10^{-9}\,s$. The phase accumulated  due to the interaction  must be equal to $\pi$. For this to happen the interaction strength must exceed $U=\frac{h}{2t} \approx 3.3*10^{-25}\,J$. The unscreened Coulomb interaction $\frac{e^2}{\epsilon r}$ ( $\epsilon\approx 12$ is dielectric constant of GaAs ) provides such a strength at the distances starting at $6,4*10^{-5}\,m$ which is about $10^4$ of  magnetic length $l_H  \approx 10^{-8}\,m$ for magnetic field $B=5\,T$. If the Fermi sea regions and various gates are at the larger distance   than the distance between the edge states in the interaction region,  using the unscreened form of Coulomb interaction  looks reasonable. Therefore needed phase accumulation is accessible in realistic structures.
 If one uses short electron wave-packets another demand appears. As short packets have wide energy distribution we must ensure that beam-splitter scattering is energy-independent within that range. The energy width of a wave-packet with duration $10^{-10}\,s$ is approximately $10\,\mu eV$. 
 This requires a short enough scattering structure (e.g. QPC)  in the beam-splitter.
 Next issue is the size of setup. If the interaction time equals to $10^{-9}\,s$ then for a drift velocity $10^4 - 10^5\,ms^{-1}$ (taken from Ref. \onlinecite{Glattli_dephasing}) interaction region length would be $ 10^{-5} - 10^{-4}\,m$. This length exceeds the size of electronic Mach-Zehnder interferometers where AB-oscillations were observed. A satisfactory length for the interaction region should be of order $10^{-6}\,m$. To achieve this length one may either produce shorter wave-packets (keeping in mind that the beam-splitters should remain energy-independent) or reduce the drift velocity via making the potential walls smoother. 
 Another issue is the accuracy of the interaction phases. First, one can tune the interaction strength with extra gates. If there is no such possibility, the setup should be fabricated with an interaction producing phase shift exceeding $\pi$. Electrons then should be sent with a delay between them to reduce the interaction time and strength. The procedure for phase control may be developed in the same style as the AB-phase adjustment described in Appendix~\ref{BS_appendix}. One should tune the bottom beam-splitter in the second MZI to transfer incoming electrons to the adjacent edge state and set the AB-phase in the first MZI to $0$. The interaction strength then should be tuned to the point where the electron sent to the first MZI comes out from the different lead. After this for better tuning one applies an AB-phase $\pi/2$. Here if the Coulomb interaction phase is equal to $\pi$ the sensitivity of one particle probabilities in the first MZI reaches its maximum. Then the same procedure should be repeated for the second interaction zone.
%It would also be good for injected electrons to have sharp lower edge of energy distribution function. With this electrons would avoid partial reflection on high potential zone. Electron-on-demand source described in Ref. \onlinecite{glattli_single,mahe}] produces exponentially decaying current which has Lorentz form in energy space with no sharp lower edge. Another proposal which is discussed in Ref. \onlinecite{Lorentz_on_demand} may fit better for our setup as it generates wave-packets exponentially decaying in energy space with sharp lower edge at Fermi level however this scheme has not been realized experimentally.   
Although it looks possible to meet the requirements listed above, the whole realization of the experiment remains very challenging. It might be therefore very interesting to realize less ambitious project with constant voltage and detection of low frequency third order current-current correlator. The latter rids of single particle detection while the former substitutes single electron sources. One can also express orbital correlators~(\ref{measproc}) in terms of current-current correlators (see Refs.~\onlinecite{chtchelkachev,beenakker2}). Third order current correlation can be measured along the lines realized in Ref.~\onlinecite{reznik} or theoretically suggested in Ref.~\onlinecite{baykos}. Still, in this case there is small hope for observation of Bell inequality violation as it requires time correlated propagation of electrons for accumulation of definite phase ($\pi$) with Coulomb interaction, which is a very rare event in the constant voltage regime - the most probable event is no phase at all since electrons usually pass interaction region separately. 
%Here we have two reasons preventing from Bell inequality violation. First, correlators are small, because probability of three particles to have interaction is small too. Second, even if interaction occurs phases are random with the most probability of small phase. 
Therefore, in the constant voltage regime Bell inequlity violation paradigm becomes impractical and one has to come back to the usual physics and compare the experiment with some specific theory. This approach does not have the main advantage of Bell inequality violation which indicates entanglement without addressing to the details of the setup construction. Instead, one has to measure irreducible third order current cross-correlators as a function of interaction strength and other parameters (e.g. AB phases and BS transparencies). From that function it should be possible to find a term originating from Coulomb interaction which would indirectly indicate entanglement. 
%To find the main source of correlation additional series of experiments with different Coulomb interaction strengths and other parameters should be performed. 
This kind of observation would also be a significant step towards the realization and detection of GHZ-state and electronic orbital states.

\section{Conclusion}
\label{conclusion}
In this article we have demonstrated, how to produce a generalized GHZ states for 3 flying qubits, 
formed as orbital electron states in the MZI-type geometries.

Our setup relies on two state of the art devices of nanophysics: a) single electron sources which 
have been demonstrated\cite{glattli_single} and characterized\cite{mahe} experimentally, 
as well as theoretically\cite{jonckheere}; 
b) Mach Zehnder interferometers, which have equally been the object of thorough experimental\cite{mach_zehnder} 
and theoretical\cite{MZ_dephasing,kovrizhin} investigations.
In our proposal, both devices should be integrated together, and moreover three MZI placed in parallel are needed to achieve
the GHZ state. While it is plausible to think that it will be challenging to build a prototype in the near future, 
we judge that it is useful for the mesoscopic physics community to be aware that advanced quantum information
protocols -- here the production of a GHZ state -- can be achieved with electrons. 

At the beginning of the paper we justified that our ideal devices generate the GHZ state. Subsequently, 
we provided a detailed explanation of the type of Bell inequalities which need to be used to 
prove unambiguously that we have generated the proper state. The problem which we encountered with the 
actual MZI setup which is needed to implement the Bell test is that it requires 3 more MZI loops than ``condensed scheme'', thus 
making the integration of the device even more challenging. Fortunately, we provided a ``condensed 
detector scheme'' were this complication can be circumvented, arguing that the operations achieved by two BS
in series can be combined with a single BS. This allowed to actually draw a ``realistic'' device
inspired from the so called ``air bridge'' technique which is used in the experiments of 
Ref. \onlinecite{mach_zehnder}.    

This work thus belongs to 
the ongoing effort called ``electron quantum optics'' where experiments and paradigms of quantum optics
are reproduced with mesoscopic physics devices. The great advantage of the present way for producing GHZ states
relies in the fact that no postselection procedure is needed here. The steps described in this paper rely on 
a unitary evolution of the initial state, and moreover, we exploit the Coulomb interaction between electron
wave packets in order to generate
the desired phase shift on the electron wave function. This particular feature departs strongly from the 
photon protocols, as photons interact weakly when travelling in vacuum. 
We thus state that the use of electron-electron interactions in single electron devices
schemes may open up new possibilities for quantum information schemes which are not 
envisionable for photons.      

Granted the present work has addressed the case of an ideal device which is free from dephasing effects.
Such phenomena are likely to affect the operation of our device as they are already known to be present in single 
MZI electronic setups.\cite{MZ_dephasing} Nevertheless our top priority is to clarify how to implement an abstract quantum information 
schemes in an actual mesoscopic device. Our analysis could be refined by taking dephasing effects into account.
The authors  plan to consider this in the future. In case of similar scheme with two MZI it is done in Ref. \onlinecite{we}      
Also we pointed out some factors that need to be accounted for in experiment and mentioned simplified scheme with constant voltage instead of on-demand electron sources.
\acknowledgements
We acknowledge financial support by the CNRS LIA agreements with Landau Institute and the RFBR Grant~\#11-02-00744-a (AAV and GBL).
T.J. and T.M. acknowledge support from grant ANR 2010 BLANC 0412 02.

\appendix

\section{GHZ maximal violation parameters}
\label{GHZ_appendix}

Here we will prove that in the tripartite case $B=4$ can be achieved only for the GHZ-state, or the states that can be reduced to GHZ with some local unitary transformations. Next we will find all possible series of measurement angles which give $B=4$ for the GHZ-state. We use the spin-based approach in this section for convenience.

First we compute the square the Bell operator~(\ref{Bell operator def}) and find conditions necessary for $\langle\hat{B}^2\rangle = 16$.
\begin{eqnarray}
\hat{B}^2 &=& 4-[\hat{\sigma}_\mathbf{a_1},\hat{\sigma}_\mathbf{a_1'}]\otimes[\hat{\sigma}_\mathbf{a_2},
\hat{\sigma}_\mathbf{a_2'}]\otimes\mathbf{1}
\nonumber\\
&~&~~-
\mathbf{1}\otimes[\hat{\sigma}_\mathbf{a_2},\hat{\sigma}_\mathbf{a_2'}]\otimes[\hat{\sigma}_\mathbf{a_3},\hat{\sigma}_\mathbf{a_3'}]
\nonumber\\
&~&~~-
[\hat{\sigma}_\mathbf{a_1},\hat{\sigma}_\mathbf{a_1'}]\otimes\mathbf{1}\otimes[\hat{\sigma}_\mathbf{a_3},\hat{\sigma}_\mathbf{a_3'}].
\end{eqnarray}
Pauli matrices ($i,j,k=x,y,z$) obey angular momentum commutation relations:
\begin{equation}
[\sigma_i,\sigma_j]=2i\epsilon^{ijk}\sigma_k,
\end{equation}
from which we obtain the commutator for a spin pointing along arbitrary vectors $\mathbf{c}$ and $\mathbf{d}$
\begin{equation}
[\sigma_\mathbf{c},\sigma_\mathbf{d}]=2i|e|\sigma_\mathbf{e},
\end{equation}
where $\mathbf{e}=\mathbf{c}\times \mathbf{d}$. For our task we will need vectors 
$\mathbf{f_1}=\mathbf{a_1}\times \mathbf{a_1'}$, $\mathbf{f_2}=\mathbf{a_2}\times \mathbf{a_2'}$, $\mathbf{f_3}=\mathbf{a_3}\times \mathbf{a_3'}$. They allow to represent $\hat{B}^2$ in the following form:
\begin{eqnarray}
\hat{B}^2&=& 4(1 + |\mathbf{f_1}||\mathbf{f_2}|\sigma_\mathbf{f_1}\otimes\sigma_\mathbf{f_2}\otimes\mathbf{1} + |\mathbf{f_2}||\mathbf{f_3}|\mathbf{1}\otimes\sigma_\mathbf{f_2}\otimes\sigma_\mathbf{f_3} \nonumber\\&~&~~+
|\mathbf{f_1}||\mathbf{f_3}|\sigma_\mathbf{f_1}\otimes\mathbf{1}\otimes\sigma_\mathbf{f_3}).
\label{B_squared}\end{eqnarray}
This operator has a maximum eigenvalue $B^2=4(1+|\mathbf{f_1}||\mathbf{f_2}|+|\mathbf{f_2}||\mathbf{f_3}|+|\mathbf{f_1}||\mathbf{f_3}|)\leq 4(1+1+1+1)=16$.

For the maximum value to be achieved we require that $|\mathbf{f_1}|=|\mathbf{f_2}|=|\mathbf{f_3}|=1$,
which implies $\mathbf{a_1}\perp \mathbf{a_1'}$, $\mathbf{a_2}\perp \mathbf{a_2'}$, $\mathbf{a_3}\perp \mathbf{a_3'}$. 
The eigenstates $\Psi$ of this value obey the relations
\begin{equation}
\begin{aligned}
\sigma_\mathbf{f_1}\otimes\sigma_\mathbf{f_2}\otimes\mathbf{1}\Psi=\Psi,\\
\mathbf{1}\otimes\sigma_\mathbf{f_2}\otimes\sigma_\mathbf{f_3}\Psi=\Psi,\\
\sigma_\mathbf{f_1}\otimes\mathbf{1}\otimes\sigma_\mathbf{f_3}\Psi=\Psi.\\
\end{aligned}
\end{equation}
It is convenient to choose a basis in each particle's space in relation with the corresponding $f$-vector. For example, $|\uparrow\rangle$ for the first spin means that $\hat{\sigma}_\mathbf{f_1}|\uparrow\rangle=|\uparrow\rangle$. In this basis the only states which give $B^2=16$ are $|\uparrow\uparrow\uparrow\rangle$ and $|\downarrow\downarrow\downarrow\rangle$, 
so in full generality we choose a superposition:
\begin{equation} 
\Psi=a|\uparrow\uparrow\uparrow\rangle+b |\downarrow\downarrow\downarrow\rangle~.
\label{superposition}
\end{equation}

We can now find which state of the form  of Eq. (\ref{superposition}) is an eigenstate of the operator $\hat{B}$ with eigenvalue $B=4$. 
Choosing the basis we associate the $\hat{z}$ axis of the first spin space with vector $\mathbf{f_1}$, and similarly for the other two spin states. 
This implies that $\mathbf{a_1}$ and $\mathbf{a_1'}$ are orthogonal vectors in the $\hat{x}-\hat{y}$ plane, and similarly for $\mathbf{a_2}$, $\mathbf{a_2'}$ and for $\mathbf{a_3}$, $\mathbf{a_3'}$.
For specificity we choose $\mathbf{a_1}\equiv \hat{x}$ and $\mathbf{a_1'} \equiv \hat{y}$. Analogously for the other spin components. Note that
\begin{eqnarray}
\sigma_\mathbf{a_1'}\sigma_\mathbf{a_2}\sigma_\mathbf{a_3}|\uparrow\uparrow\uparrow\rangle
&=&\sigma_\mathbf{a_1}\sigma_{\mathbf{a_2'}}\sigma_\mathbf{a_3}|\uparrow\uparrow\uparrow\rangle
=\sigma_\mathbf{a_1}\sigma_\mathbf{a_2}\sigma_\mathbf{a_3'}|\uparrow\uparrow\uparrow\rangle\nonumber\\
&=&-\sigma_\mathbf{a_1'}\sigma_\mathbf{a_2'}\sigma_\mathbf{a_3'}|\uparrow\uparrow\uparrow\rangle = i|\downarrow\downarrow\downarrow\rangle, \nonumber\\
\sigma_\mathbf{a_1'}\sigma_\mathbf{a_2}\sigma_\mathbf{a_3}|\downarrow\downarrow\downarrow\rangle 
&=&\sigma_\mathbf{a_1}\sigma_\mathbf{a_2'}\sigma_\mathbf{a_3}|\downarrow\downarrow\downarrow\rangle
=\sigma_\mathbf{a_1}\sigma_\mathbf{a_2}\sigma_\mathbf{a_3'}|\downarrow\downarrow\downarrow\rangle\nonumber\\
&=&-\sigma_\mathbf{a_1'}\sigma_\mathbf{a_2'}\sigma_\mathbf{a_3'}|\downarrow\downarrow\downarrow\rangle = -i|\uparrow\uparrow\uparrow\rangle,
\end{eqnarray}
thus implying that
\begin{equation}
\hat{B}(a|\uparrow\uparrow\uparrow\rangle+b|\downarrow\downarrow\downarrow\rangle)=-4ib|\uparrow\uparrow\uparrow\rangle+ia|\downarrow\downarrow\downarrow\rangle).
\end{equation}
From this we easily confirm that when $b=ia$ we have the desired eigenstate with $B=4$
\begin{equation}
\Psi = \frac{1}{\sqrt{2}}(|\uparrow\uparrow\uparrow\rangle+i|\downarrow\downarrow\downarrow\rangle).
\end{equation}
Also, note that the orthogonal state $\Psi' = \frac{1}{\sqrt{2}}(|\uparrow\uparrow\uparrow\rangle-i|\downarrow\downarrow\downarrow\rangle)$ has an eigenvalue $B=-4$. Both states are equivalent to the GHZ state up to local unitary transformations.
In conclusion we have shown that each state which can violate the Bell-type inequality maximally is a GHZ-state up to trivial transformations. 

Now we proceed with the reverse problem. Starting from the GHZ-state, let us find all sets of vectors 
$\mathbf{a_1}$,  $\mathbf{a_2}$, $\mathbf{a_3}$, 
$\mathbf{a_1'}$,  $\mathbf{a_2'}$, $\mathbf{a_3'}$, which yield $\bar{B}=4$. 
We consider a spin correlator $\langle\hat{\sigma}_\mathbf{n_1}\otimes\hat{\sigma}_\mathbf{n_2}\otimes\hat{\sigma}_\mathbf{n_3}\rangle$ for arbitrary unitary vectors $\mathbf{n_1}$, $\mathbf{n_2}$, $\mathbf{n_3}$ with corresponding spherical coordinates $\theta_\mathbf{n_1}, \phi_\mathbf{n_1}, \theta_\mathbf{n_2}, \phi_\mathbf{n_2}, \theta_\mathbf{n_3}, \phi_\mathbf{n_3}$.

The operator for the spin projection on direction $\mathbf{n}$ is then
\begin{equation}
\hat{\sigma}_\mathbf{n} = n_x\hat{\sigma}_x + n_y\hat{\sigma}_y + n_z\hat{\sigma}_z=
\begin{pmatrix}
\cos{\theta_\mathbf{n}} & \sin{\theta_\mathbf{n}}e^{-i\phi_\mathbf{n}}\\
\sin{\theta_{\mathbf{n}}}e^{i\phi_\mathbf{n}} & -\cos{\theta_\mathbf{n}}\\
\end{pmatrix}.
\end{equation}
This allows us to find
\begin{multline}
\hat{\sigma}_\mathbf{n_1}\otimes\hat{\sigma}_\mathbf{n_2}\otimes\hat{\sigma}_\mathbf{n_3}|\uparrow\uparrow\uparrow\rangle =\\ \cos{\theta_\mathbf{n_1}}\cos{\theta_\mathbf{n_2}}\cos{\theta_\mathbf{n_3}}|\uparrow\uparrow\uparrow\rangle +\\   
 \sin{\theta_\mathbf{n_1}}\sin{\theta_\mathbf{n_2}}\sin{\theta_\mathbf{n_3}}e^{i(\phi_\mathbf{n_1}+\phi_\mathbf{n_2}+\phi_\mathbf{n_3})}|\downarrow\downarrow\downarrow\rangle+ ...\\
\end{multline}
\begin{multline}
\hat{\sigma}_\mathbf{n_1}\otimes\hat{\sigma}_\mathbf{n_2}\otimes\hat{\sigma}_\mathbf{n_3}|\downarrow\downarrow\downarrow\rangle =\\ \sin{\theta_\mathbf{n_1}}\sin{\theta_\mathbf{n_2}}\sin{\theta_\mathbf{n_3}}e^{-i(\phi_\mathbf{n_1}+\phi_\mathbf{n_2}+\phi_\mathbf{n_3})}|\uparrow\uparrow\uparrow\rangle -\\ 
\cos{\theta_\mathbf{n_1}}\cos{\theta_\mathbf{n_2}}\cos{\theta_\mathbf{n_3}}|\downarrow\downarrow\downarrow\rangle + ...\\
\end{multline}
and the average of the spin correlator in the GHZ state is then:
\begin{equation}
\langle\hat{\sigma}_\mathbf{n_1}\otimes\hat{\sigma}_\mathbf{n_2}\otimes\hat{\sigma}_\mathbf{n_3}\rangle = %\frac{e^{i(\phi_m+\phi_n+\phi_k)}+e^{-i(\phi_m+\phi_n+\phi_k)}}{2} = 
\sin{\theta_\mathbf{n_1}}\sin{\theta_\mathbf{n_2}}\sin{\theta_\mathbf{n_3}}\cos(\phi_\mathbf{n_1}+\phi_\mathbf{n_1}+\phi_\mathbf{n_3}).
\end{equation}
The simple formula which we have obtained is a justification of the spin-based approach to Bell type inequalities.
The tripartite case is simpler than the bipartite state when the question comes to finding the specific 
angles giving Tsirel'son bound, since here all the correlators should have unit absolute value. This implies 
\begin{equation}
\sin{\theta_\mathbf{n_1}} = \sin{\theta_\mathbf{n_2}} = \sin{\theta_\mathbf{n_3}} = 1,
\end{equation}   
and thus we should consider only vectors of the $\hat{x}-\hat{y}$ plane as candidates for $B=4$.
The  polar angles of these vectors $\mathbf{a_1}$, $\mathbf{a_2}$, $\mathbf{a_3}$,$\mathbf{a_1'}$, $\mathbf{a_2'}$, $\mathbf{a_3'}$ are defined by the conditions:
\begin{eqnarray}
\langle\hat{\sigma}_\mathbf{a_1'}\otimes\hat{\sigma}_\mathbf{a_2}\otimes\hat{\sigma}_\mathbf{a_3}\rangle &=& 1, \nonumber\\
\langle\hat{\sigma}_\mathbf{a_1}\otimes\hat{\sigma}_\mathbf{a_2'}\otimes\hat{\sigma}_\mathbf{a_3}\rangle
& =& 1, \nonumber\\
\langle\hat{\sigma}_\mathbf{a_1}\otimes\hat{\sigma}_\mathbf{a_2}\otimes\hat{\sigma}_\mathbf{a_3'}\rangle 
&=& 1, \nonumber\\
\langle\hat{\sigma}_\mathbf{a_1'}\otimes\hat{\sigma}_\mathbf{a_2'}\otimes\hat{\sigma}_\mathbf{a_3'}\rangle 
&=& -1,
\end{eqnarray}
which in terms of polar angles reads:
\begin{eqnarray}
\cos(\phi_\mathbf{a_1'} + \phi_\mathbf{a_2} + \phi_\mathbf{a_3}) &=& 1,\nonumber\\
\cos(\phi_\mathbf{a_1} + \phi_\mathbf{a_2'} + \phi_\mathbf{a_3}) &=& 1,\nonumber\\
\cos(\phi_\mathbf{a_1} + \phi_\mathbf{a_2} + \phi_\mathbf{a_3'}) &=& 1, \nonumber\\
\cos(\phi_\mathbf{a_1'} + \phi_\mathbf{a_2'} + \phi_\mathbf{a_3'}) &=& -1. 
\end{eqnarray}

This system of equations has the solution:
\begin{equation}
\phi_\mathbf{a_1}+\phi_\mathbf{a_2}+\phi_\mathbf{a_3} =\pm\frac{\pi}{2}.\label{angles}
\end{equation}
with $\phi_\mathbf{a_i'} = \phi_\mathbf{a_i} \mp \pi/2$, where $i=1, 2, 3$.
All the vectors lye in the $\hat{x}-\hat{y}$ plane.

This concludes the definition of the angles giving $B=4$ for the GHZ-state.

\section{Beam-splitter parametrization}
\label{BS_appendix}

One of essential parts of the device is the BS. The BS is a 4-arm scatterer with a
special type of scattering matrix which allows to split the beam into two parts without any reflection.
Generally the scattering matrix of 4-arm splitter looks like 
\begin{equation}
S=\begin{pmatrix}
r_{11} & t_{21} & t_{31} & t_{41} \\
t_{12} & r_{22} & t_{32} & t_{42} \\
t_{13} & t_{23} & r_{33} & t_{43} \\
t_{14} & t_{24} & t_{34} & r_{44}
\end{pmatrix}.
\end{equation} 
In this matrix the $t_{ij}$ parameters refer to the transmission amplitudes from arm $j$ to arm $i$, and $r_{ii}$ is the 
reflection amplitude in arm $i$.
This matrix obeys only a unitarity condition. If time-reversal invariance is added a
new condition appears $t_{ij} = t_{ji}$. The beam-splitter scattering matrix is
\begin{equation}
S=\begin{pmatrix}
0 & 0 & t_{31} & t_{41} \\
0 & 0 & t_{32} & t_{42} \\
t_{31} & t_{32} & 0 & 0 \\
t_{41} & t_{42} & 0 & 0
\end{pmatrix}.
\end{equation}
For these parameters the following system of equations is valid
\begin{eqnarray}
T_{31} + T_{41} &=& 1,\nonumber\\
T_{31} + T_{32} &=& 1,\nonumber\\
T_{32} + T_{42} &=& 1,\nonumber\\
t_{31}t^*_{32} + t_{41}t^*_{42} &=& 0.
\end{eqnarray}
Where $T_{ij} = |t_{ij}|^2.$ For such an equation set it is rather easy to
find a simple parametrization. Let us assume that
$t_{31}=\cos{\theta}e^{i\phi_{31}},
t_{41}=\sin{\theta}e^{i\phi_{41}},
t_{32}=\sin{\theta}e^{i\phi_{32}},
t_{42}=\cos{\theta}e^{i\phi_{42}}.$
Then we substitute all the t-values into last equation and get:
\begin{equation}
\cos{\theta}\sin{\theta}e^{i\phi_{31}-i\phi_{32}}+\sin{\theta}\cos{\theta}e^{i\phi_{41}-i\phi_{42}} = 0.
\end{equation}
For this equation to be valid must be
\begin{equation}
\phi_{31}-\phi_{32}-\phi_{41}+\phi_{42}=\pi+2\pi n, n\in\mathbb{Z}.\label{phase cond}
\end{equation}
In the present work we focus on symmetrical beam-splitters for simplicity. For  a symmetrical beam-splitter
\begin{equation}
t_{31}=t_{42},\qquad t_{32}=t_{41}.
\end{equation}
so the phases are $\phi_{31}=\phi_{42}, \phi_{32}=\phi_{41}$ and Eq.~(\ref{phase cond}) simplifies to:
\begin{equation}
\phi_{31}-\phi_{32}=\pi/2+\pi n, n\in\mathbb{Z}.
\end{equation}

One of the possible solutions is
\begin{eqnarray}
\phi_{31} = 0,     &t_{31} = t_{42} = \cos{\theta},\nonumber\\
\phi_{32} = \pi/2, &t_{32} = t_{41} = i\sin{\theta}.
\end{eqnarray}

This parametrization will be used in further calculations.

%The difference produced by various parametrisations can be killed with additional phase difference between left and right paths in each MZI (in %case of different symmetrical parametrizations it equals to $\pi$). This phase difference can be introduced for example by external magnetic %field (via Aharonov-Bohm effect).

We are interested in the properties of electron transport from bottom to top, so we will need only the values of $t_{31}, t_{32}, t_{41}, t_{42}$. Now we can introduce the transfer matrix:
\begin{equation}
T = \begin{pmatrix}
t_{RR} & t_{LR}\\ \label{transfer-matr}
t_{RL} & t_{LL}
\end{pmatrix} = \begin{pmatrix}
i\sin{\theta} & \cos{\theta}\\
\cos{\theta} & i\sin{\theta}
\end{pmatrix}~.
\end{equation}
Note that we have changed indexes because of the reflectionless nature of the BS: $t_{RR} = t_{32}, t_{RL} = t_{42}, t_{LR} = t_{31}, t_{LL} = t_{31}$.
In order to find the state of particles after the BS one must perform the following substitution with wave-functions before the beam-splitter:
\begin{eqnarray}
\psi_R & \rightarrow & \cos{\theta}\psi_L + i\sin{\theta}\psi_R,\nonumber\\
\psi_L & \rightarrow & \cos{\theta}\psi_R + i\sin{\theta}\psi_L.
\end{eqnarray}

Since we have chosen a specific parametrisation of the BS, we must confirm that if the BS used in the experiment has a transfer matrix which is different from the one in Eq. (\ref{transfer-matr}) it will not spoil the whole experiment.

%add more whistles about different parametrisations

We consider again the phase relation in the general case:
\begin{equation}
\phi_{31}-\phi_{32}-\phi_{41}+\phi_{42}=\pi+2\pi n, n\in\mathbb{Z}
\end{equation}

Each phase can be presented as:
\begin{equation}
\begin{aligned}
\phi_{31} & = \tilde{\phi}_{31} + 2\pi n \equiv \tilde{\phi}_{31},\\
\phi_{32} & = \pi/2 + \tilde{\phi}_{32},\\
\phi_{41} & = \pi/2 + \tilde{\phi}_{41},\\ 
\phi_{42} & = \tilde{\phi}_{42},
\end{aligned}
\end{equation}
where the phases $\tilde{\phi}_{31},\tilde{\phi}_{32},\tilde{\phi}_{41},\tilde{\phi}_{42}$ represent the deviation from the chosen parametrisation. These obey the relation:
\begin{equation}
\tilde{\phi}_{31}-\tilde{\phi}_{32}-\tilde{\phi}_{41}+\tilde{\phi}_{42} = 0.
\end{equation}

All solutions of this equation are taken into account by the following representation:
\begin{equation}
\begin{aligned}
\tilde{\phi}_{31} &= \phi_3,\\
\tilde{\phi}_{32} &= \phi_3 + \phi_2,\\
\tilde{\phi}_{41} &= \phi_4,\\
\tilde{\phi}_{42} &= \phi_4 + \phi_2.\\
\end{aligned}
\end{equation}
where $\phi_2,\phi_3,\phi_4$ are some real values.

\begin{figure}[t]
\centerline{\includegraphics[width=2.5cm]{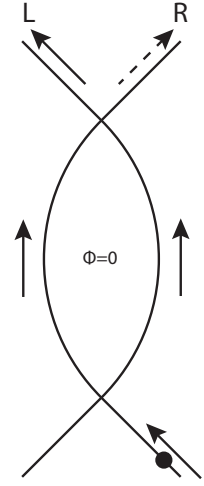}}
\caption{For $\Phi=0$ electron goes to left arm.}\label{1MZI}
\end{figure}

This answer has an explicit physical meaning. All beam-splitters have the same transfer matrix up to the phase accumulation in each channel. This phase accumulation can affect the phase difference in the left and the right paths of the MZI. In order to restore the original interference pattern we need to adjust the applied phase difference. During the experiment it means choosing a reference point with $\Phi=0$ for given MZI. This could be achieved using the fact that at $\Phi=0 $ an electron injected in the right arm of the MZI has zero probability to be detected in the right arm~(fig.~\ref{1MZI}). Yet better tuning can be achieved near $\pi/2$. Here the probabilities for each arm are equal and the difference between the probabilities has a maximum derivative with respect to the phase.

{\it Beam-splitter with phase}.

In our MZI the electrons have two possible paths to travel. If there is a difference in the accumulated phase between two paths, it will affect the interference between them in a subsequent beam-splitter. It is convenient to account for this phase difference (which can be generated e.g. by applying a magnetic flux through the MZI loop) in the transfer matrix of the BS.
If the phase lengths of the right and left the paths are $\phi_L$ and $\phi_R$, then the transfer matrix of the beam-splitter takes the form:
\begin{equation}
T = \begin{pmatrix}
i\sin{\theta}e^{i\phi_R} & \cos{\theta}e^{i\phi_L}\\
\cos{\theta}e^{i\phi_R} & i\sin{\theta}e^{i\phi_L}\\
\end{pmatrix}.\label{transfer matr}
\end{equation}
Actually the final results depend only on the phase difference of the two paths $\Phi = \phi_L - \phi_R$
so the transfer matrix can be multiplied by any phase factor without affecting the final results. 
We can represent the transfer matrix in a form similar to the spin rotation operator
\begin{equation}
T = \begin{pmatrix}
\sin{\theta}e^{-i\Phi/2} & -i\cos{\theta}e^{i\Phi/2}\\
-i\cos{\theta}e^{-i\Phi/2} & \sin{\theta}e^{i\Phi/2}\\
\end{pmatrix} = e^{-i\sigma_x\frac{(\pi-2\theta)}{2}}e^{-i\sigma_z\frac{\Phi}{2}}.
\end{equation} 

Since the measurement procedure in the condensed scheme includes changing $\theta,\Phi$ and measuring $\langle\sigma_z\otimes\sigma_z\otimes\sigma_z\rangle$ at level ``(lv2')''~(Fig.~\ref{setup(f)}) it is convenient to define a measurement operator with respect to level ``lv2'' of the scheme, where the electron state is invariant during the measurement procedure.

As we see the operator is a sum of two rotations, first a rotation of angle $\Phi$ around the $z$ axis, second another rotation of angle $\pi-2\theta$ around the $x$ axis. The rotation around x-axis corresponds to the action of the beam-splitter itself, while the
rotation around $z$ corresponds to an Aharonov-Bohm phase accumulation. This whole rotation transforms a vector
\begin{equation}
\mathbf{n} = \begin{pmatrix}
\sin{2\theta}\sin{\Phi}\\
\sin{2\theta}\cos{\Phi}\\
-\cos{2\theta}
\end{pmatrix}\label{character vector},
\end{equation}
unto the $z$ axis. As the vector thus defined is in one-to-one correspondence with the pair of BS parameters $\theta,\Phi$ we call it a characteristic vector of the BS. Rotation means that
\begin{equation}
T|\mathbf{n}\rangle = U_{\mathbf{n}\rightarrow z}|\mathbf{n}\rangle = |\uparrow\rangle
\end{equation} 
The projection of the rotated pseudospin onto $z$ can be shown to be equal to the projection of the initial pseudospin onto $\mathbf{n}$.
\begin{equation}
\langle U_{\mathbf{n}\rightarrow z}\Psi|\sigma_z|U_{\mathbf{n}\rightarrow z}\Psi\rangle=
\langle\Psi|U^+_{\mathbf{n}\rightarrow z}\sigma_zU_{\mathbf{n}\rightarrow z}|\Psi\rangle = 
\langle\Psi|\sigma_\mathbf{n}|\Psi\rangle.
\end{equation}

For three-particle spin-correlators the same relation holds
%\begin{equation}
%\langle U_{\mathbf{c_1}\rightarrow z} \otimes U_{\mathbf{c_2}\rightarrow z} \otimes U_{\mathbf{c_3}\rightarrow z} \Psi |\sigma_z \otimes \sigma_z \otimes \sigma_z |U_{\mathbf{c_1} \rightarrow z} \otimes U_{\mathbf{c_2} \rightarrow z} \otimes U_{\mathbf{c_3} \rightarrow z} \Psi \rangle = 
%\langle \Psi | U^+_{\mathbf{c_1}\rightarrow z}\sigma_z U_{\mathbf{c_1} \rightarrow z}  \otimes U^+_{\mathbf{c_2}\rightarrow z}\sigma_zU_{\mathbf{c_2} \rightarrow z} \otimes U^+_{\mathbf{c_3}\rightarrow z}\sigma_zU_{\mathbf{c_3}\rightarrow z}|\Psi\rangle = 
%\langle\Psi|\sigma_\mathbf{c_1}\otimes\sigma_\mathbf{c_2}\otimes\sigma_\mathbf{c_3}|\Psi\rangle 
%\end{equation} 

\begin{multline}
\langle T_1 \otimes T_2 \otimes T_3 \Psi |\sigma_z \otimes \sigma_z \otimes \sigma_z |T_1 \otimes T_2 \otimes T_3 \Psi \rangle =\\ 
\langle \Psi | T^+_1\sigma_z T_1  \otimes T^+_2\sigma_zT_2 \otimes T^+_3\sigma_zT_3|\Psi\rangle = \\
\langle\Psi|\sigma_\mathbf{n_1}\otimes\sigma_\mathbf{n_2}\otimes\sigma_\mathbf{n_3}|\Psi\rangle 
\end{multline}  

So now we have found that the measured correlator with respect to level ``lv2'' is $\langle\sigma_\mathbf{n_1}\otimes\sigma_\mathbf{n_2}\otimes\sigma_\mathbf{n_3}\rangle$.

\section{Pseudospin analogy}
\label{pseudospin analogy}

Previously in appendix \ref{GHZ_appendix} we have defined angles yielding $B=4$ for three-spin GHZ-state. To obtain the parameters of the actual setup~(Fig.~\ref{setup(f)}) we need to ``translate'' the spin angles into BS parameters. This chapter contains the derivation of the ``translation'' rules.

We start with the logic of the GHZ-experiment described in Sec. \ref{GHZ_section}. 
We need three values $x_1,x_2,x_3$ corresponding to separate measurements with results within the interval $[-1;1]$. In the spin case, the spin projection measurements serve this purpose. In our setup we can detect the particle in the left or the right arm. So one of possible assignments is $x=-1$ for the particle detected in left arm and $x=1$ for particle detected in right arm. Then the Bell correlator is

\begin{eqnarray}
\langle x_1x_2x_3\rangle &=& P_{RRR} + P_{RLL} + P_{LRL} + P_{LLR}\nonumber\\
~~~&~&- P_{LLL} - P_{LRR} - P_{RLR} - P_{RRL}.\label{corel}
\end{eqnarray}

This correlator measured on level ``lv3'' of the setup~(Fig~.\ref{setup(f)}) corresponds to the  $\langle\hat{\sigma}_z\otimes\hat{\sigma}_z\otimes\hat{\sigma}_z\rangle$ correlator of pseudospin. 

Due to the rotation between ``(lv2')'' and level ``lv3'' the correlator of Eq.~(\ref{corel}) corresponds to $\langle\hat{\sigma}_\mathbf{n_1}\otimes\hat{\sigma}_\mathbf{n_2}\otimes\hat{\sigma}_\mathbf{n_3}\rangle$
measured on the initial pseudospin state at ``(lv2')'' which is GHZ. The vectors $\mathbf{n_1},\mathbf{n_2},\mathbf{n_3}$ 
are characteristic vectors of the corresponding beam-splitters~(\ref{character vector}).

Now we calculate BS matrices corresponding to the angles giving maximum violation~(\ref{angles}).
All the vectors are in the $\hat{x}-\hat{y}$ plane, thus $\theta = \pi/4$ for all of them. For the 
vector with the polar angle $\phi$ corresponding $\Phi =\alpha = \pi/2 - \phi $ and the BS transfer matrix is

\begin{equation}
T(\alpha) = \begin{pmatrix}
\frac{1}{\sqrt{2}}e^{-i\alpha/2} & -\frac{i}{\sqrt{2}}e^{i\alpha/2}\\
-\frac{i}{\sqrt{2}}e^{-i\alpha/2} & \frac{1}{\sqrt{2}}e^{i\alpha/2}\\
\end{pmatrix} \equiv \begin{pmatrix}
\frac{i}{\sqrt{2}} & \frac{1}{\sqrt{2}}e^{i\alpha}\\
\frac{1}{\sqrt{2}} & \frac{i}{\sqrt{2}}e^{i\alpha}\\
\end{pmatrix},
\end{equation}
where equivalence means that the overall phase does not change the observable outcomes. $\alpha$ parameters then should obey the restrictions: 
\begin{equation}
\begin{aligned}
\alpha_\mathbf{a_1} + \alpha_\mathbf{a_2} +\alpha_\mathbf{a_3} = \pi/2 \pm \pi/2,\\
\alpha_\mathbf{a'_i} = \alpha_\mathbf{a_i} \pm  \pi/2,\\
\end{aligned}
\end{equation}
for $i=1,2,3$.

\section{Settings giving maximum violation of Bell inequality.}
\label{settings_appendix}

In appendix \ref{pseudospin analogy} we have defined the properties of BS between levels ``(lv2')'' and ``lv3'' of Fig.~\ref{setup(f)}. The next step lies in defining the parameters of BS in the condensed scheme. For this purpose we use the properties of unitary transformation made between levels ``lv2'' and ``(lv2')'' of the initial setup~(\ref{transformation}).

The problem is simplified by the fact that all BS between levels ``(lv2')'' and ``lv3'' are of the same type
\begin{equation}
T(\alpha) = \begin{pmatrix}
\frac{i}{\sqrt{2}} & \frac{1}{\sqrt{2}}e^{i\alpha}\\
\frac{1}{\sqrt{2}} & \frac{i}{\sqrt{2}}e^{i\alpha}\\
\end{pmatrix}
\end{equation}

\textit{First and third MZI.}

Since $U_3 = U_1$ all the results are valid for the third MZI as well.
The first MZI chain makes two consequent unitary transformations, the first one is $U_1 = \frac{1}{\sqrt{2}}\begin{pmatrix}
-i & 1\\
1 & -i\\
\end{pmatrix},$ next is $T(\alpha) = \frac{1}{\sqrt{2}}\begin{pmatrix}
i & e^{i\alpha}\\
1 & ie^{i\alpha}\\
\end{pmatrix}$.
The product of these is unitary 
\begin{equation}
T_1(\alpha) = T(\alpha)U_1 = \frac{1}{2}\begin{pmatrix}
1+e^{i\alpha} & i-ie^{i\alpha}\\
-i+ie^{i\alpha} & 1+e^{i\alpha}\\
\end{pmatrix}
\end{equation}
Shifting the whole matrix phase by $-\alpha/2$ we get following expression
\begin{equation}
T_1(\Phi) = \begin{pmatrix}
\cos(\alpha/2) & \sin(\alpha/2)\\
-\sin(\alpha/2) & \cos(\alpha/2)\\
\end{pmatrix} \label{matrix1}
\end{equation}
The next step is to find from the matrix the beam-splitter the parameters $\theta$, $\Phi$. 
We encounter the  problem that the matrix of Eq.~(\ref{matrix1}) cannot represent the transfer-matrix of any BS. 
The relation $\arg(t_{RR}/t_{RL}) = \frac{\pi}{2}$ which is universal for any representation of the transfer matrix (see Eq.~(\ref{transfer matr})) is not valid for Eq.~(\ref{matrix1}). The BS transfer matrix should not necessarily be equal to $T_1(\Phi)$ but it should lead to the same three-particle probabilities. 
%$P_{ijk}$, where $i,j,k$ can take values $R, L$. 
This happens if transfer matrix of beam-splitter equals to:
\begin{equation}
T_1(\Phi,\phi_1,\phi_2) = \begin{pmatrix}
\cos(\alpha/2)e^{i\phi_1} & \sin(\alpha/2)e^{i\phi_1}\\
-\sin(\alpha/2)e^{i\phi_2} & \cos(\alpha/2)e^{i\phi_2}\\
\end{pmatrix}
\end{equation}
where $\phi_1$ and $\phi_2$ are some phase shifts. They do not contribute to tripartite probabilities while this matrix with properly chosen phase shifts can be transfer matrix of BS. 

If $\tan(\alpha/2) > 0$ then $\phi_1=\pi/2$, $\phi_2=\pi$ and  
\begin{equation}
\begin{aligned}
\theta = \frac{\pi-\alpha}{2},\\
\Phi = \frac{\pi}{2}.\\
\end{aligned}
\end{equation} 

If $\tan(\alpha/2) < 0$ then $\phi_1=\pi/2$, $\phi_2=0$ and
\begin{equation}
\begin{aligned}
\theta = \frac{\pi+\alpha}{2},\\
\Phi = -\frac{\pi}{2}.\\
\end{aligned}
\end{equation}
    
Both formulas can be unified as:
\begin{equation}
\begin{aligned}
\theta = \frac{\pi-\sign{\tan(\alpha/2)}\alpha}{2},\\
\Phi = \sign{\tan(\alpha/2)}\frac{\pi}{2}.\\
\end{aligned}
\end{equation}

If $\alpha \in [-\pi;\pi]$ last formula is simplified
\begin{equation}
\begin{aligned}\label{BS1params}
\theta = \frac{\pi-|\alpha|}{2},\\
\Phi = \frac{\pi|\alpha|}{2\alpha}.\\
\end{aligned}
\end{equation}

\textit{Second MZI.}

For second MZI everything is simpler since $U_2$ is a diagonal matrix. 
\begin{equation}
T_2(\alpha) = T(\alpha)U_2 = \frac{1}{\sqrt{2}}\begin{pmatrix}
-1 & e^{i\alpha}\\
i & ie^{i\alpha}\\
\end{pmatrix}\equiv \frac{1}{\sqrt{2}}\begin{pmatrix}
i & e^{i(\alpha-\pi/2)}\\
1 & ie^{i(\alpha-\pi/2)}\\
\end{pmatrix}
\end{equation} 
This matrix corresponds to a BS with parameters $\theta=\pi/4$ and $\Phi=\alpha-\pi/2$.

Finally we can introduce the algorithm for the generation of the measurement settings resulting in $B=4$. 
\begin{enumerate}
\item Choose angles $\alpha_\mathbf{a_1},\alpha_\mathbf{a_2}, \alpha_\mathbf{a_3}$ obeying relations 
\begin{equation}
\begin{aligned}
\alpha_\mathbf{a_1} + \alpha_\mathbf{a_2}+ \alpha_\mathbf{a_3}=\pi/2 \pm \pi/2,\\
\alpha_\mathbf{a'_i} = \alpha_\mathbf{a_i} \pm \pi/2,\\
\end{aligned}
\end{equation}
where $i=1,2,3$.
\item With necessary $2\pi$ shifts make $\alpha_\mathbf{a'_i},\alpha_\mathbf{a_i}$ be in $[-\pi;\pi]$; 
\item Calculate BS parameters
\begin{equation}
\begin{aligned}
\theta_{1,3} &= \frac{\pi-|\alpha_\mathbf{a_{1,3}}|}{2}, \qquad 
&\theta'_{1,3} = \frac{\pi-|\alpha_\mathbf{a'_{1,3}}|}{2},& \\
\Phi_{1,3} &= \frac{\pi|\alpha_\mathbf{a_{1,3}}|}{2\alpha_\mathbf{a_{1,3}}},   
\qquad &\Phi'_{1,3} = \frac{\pi|\alpha_\mathbf{a'_{1,3}}|}{2\alpha_\mathbf{a'_{1,3}}},& \\ 
\theta_2 &= \pi/4, \qquad &\theta'_2 = \pi/4, & \\
\Phi_2 &= \alpha_\mathbf{a_2}-\pi/2,     \qquad &\Phi'_2 = \alpha_\mathbf{a'_2}-\pi/2, & \\
\end{aligned}
\end{equation}
\end{enumerate}

From the last equation we can extract following common features of these measurement sets.
\begin{itemize}
\item In the first and the third MZI the AB phase can take values $\pm\pi/2$;
\item In the second MZI the transparency of the BS is not changing during the experiment;
\end{itemize}

Among measurement schemes there exist ones where transparencies of all BS do not change during experiment. For example, choosing 
\begin{equation}
\begin{aligned}
\alpha_\mathbf{a_1} &= -3\pi/4, \qquad \alpha_\mathbf{a'_1} &=& 3\pi/4,  \\
\alpha_\mathbf{a_2} &= -\pi/2,  \qquad \alpha_\mathbf{a'_2} &=& \pi,     \\
\alpha_\mathbf{a_3} &= -3\pi/4, \qquad \alpha_\mathbf{a'_3} &=& 3\pi/4,  \\
\end{aligned}
\end{equation}

we get

\begin{equation}
\begin{aligned}
\theta_1 &= \pi/8, \qquad &\theta'_1 &=& \pi/8,\\
\Phi_1   &= -\pi/2,\qquad &\Phi'_1   &=& \pi/2,\\ 
\theta_2 &= \pi/4, \qquad &\theta'_2 &=& \pi/4,\\
\Phi_2   &= \pi,   \qquad &\Phi'_2   &=& \pi/2,\\
\theta_3 &= \pi/8, \qquad &\theta'_3 &=& \pi/8,\\
\Phi_3   &= -\pi/2,\qquad &\Phi'_3   &=& \pi/2,\\
\end{aligned}
\end{equation}

In conclusion, we found the general parametrization for measurement settings giving $B=4$ in a regular MZ setup, and described a special case where we get a maximum violation only by adjusting AB phases.

\section{Coulomb interaction.}
\label{Coulomb interaction.}

Here we solve the two-particle problem for the propagation of two electrons in neighbouring channels between the two beam splitters. 
We consider two chiral edge-channels of the IQHE. The two particles propagating in different channels interact 
electrostatically within a specified interaction region. The potential of interaction $U(x_1,x_2)$ depends on the 
position of both particles. The energy spectrum is assumed to be linear and the drift velocity of the electrons in each channel is the same $v_D$. 
We consider the Hilbert space of two electron wave functions such that 
$\int \Psi(x_1,x_2) e^{-ik_1x_1-ik_2x_2}dx_1dx_2 = 0$ if $k_1$ or $k_2$ is negative.
The Shr\"{o}dinger equation for this two particle problem reads:
\begin{multline}
i\frac{\partial}{\partial t}\Psi(x_1,x_2,t) = (-iv_D\frac{\partial}{\partial x_1}-iv_D\frac{\partial}{\partial x_2})\Psi(x_1,x_2,t) + \\
U(x_1,x_2)\Psi(x_1,x_2,t).
\end{multline} 
Let's make a substitution $a_\alpha=x_\alpha-v_Ft$. Then $\Psi(x_1,x_2,t)=\Phi(a_1,a_2,t)$. Here 
\begin{equation}
\frac{\partial}{\partial t} = \frac{\partial}{\partial t} + v_D\frac{\partial}{\partial x_1} + v_D\frac{\partial}{\partial x_2}
\end{equation} 
and this allows to rewrite the equation
\begin{equation}
\frac{\partial}{\partial t}\Phi = -iU(a_1+v_Dt,a_2+v_Dt)\Phi. 
\end{equation}
The solution is 
\begin{equation}
\Phi(a_1,a_2,t) = \Phi(a_1,a_2,0)e^{-i\int_0^tU(a_1+v_D\tau ,a_2+v_D\tau)d\tau}.
\end{equation}
In terms of the initial variables
\begin{multline}
\Psi(x_1,x_2,t) = \Psi_0(x_1-v_Dt,x_2-v_Dt)\times\\
e^{-i\int_0^t U(x_1+v_D(\tau-t),x_2+v_D(\tau-t))d\tau}.
\end{multline}

The phase accumulated during the interaction is position dependent and describes the energy exchange and the  deformation of the wave-packets. Here the condition for simple phase accumulation is that the size of the wave-packets is small. In this case the position dependence of the phase should not manifest itself. The solution agrees well with the derivation of Ref. \onlinecite{another 2 MZ} except for the fact the interaction kernel differs (we used a kernel of general form $U(x_1,x_2)$).


\begin{thebibliography}{99}
\bibitem{EPR} A. Einstein, B.Podolsky and N. Rosen, Phys. Rev. \textbf{47}, 777 (1935).
\bibitem{Schrodinger cat} E. Schr\"{o}dinger, Die Naturwissenschaften \textbf{23}, 807(1935).
\bibitem{Bell original} J. S. Bell, Phys. \textbf{1}, 195 (1964).
\bibitem{CHSH}  J. F. Clauser, M. A. Horne, A. Shimony and R. A. Holt, Phys. Rev. Lett. \textbf{23}, 880 (1969)
\bibitem{2_photons} A. Aspect, P. Grangier and G. Roger, Phys. Rev. Lett. {\bf 47}, 460 (1981).
\bibitem{lesovik_martin_blatter} G. Lesovik, T. Martin and G. Blatter,
European Physical J. B {\bf 24}, 287 (2001). 
\bibitem{loss} Patrik Recher, Eugene V. Sukhorukov and Daniel Loss
Phys. Rev. B {\bf 63}, 165314 (2001).
\bibitem{beenakker_2particle} C. W. J. Beenakker,  C. Emary, M. Kindermann, and J. L. van Velsen, Phys. Rev. Lett. {\bf 91}, 147901 (2003).
\bibitem{buttiker_2mz} P. Samuelsson, E. V. Sukhorukov, and M. B\"uttiker, Phys. Rev. Lett. {\bf 91}, 157002 (2003); {\it ibid.} {\bf 92}, 026805 (2004).
\bibitem{chtchelkachev} N. M. Chtchelkatchev, G. Blatter, G. B. Lesovik and T.
Martin, Phys. Rev. B {\bf 66}, 161320 (2002).
\bibitem{lebedevlesovik} A. V. Lebedev, G. B. Lesovik and G. Blatter, Phys. Rev. B {\bf 71}, 045306 (2005).
\bibitem{sauret} O. Sauret, T. Martin and D. Feinberg, Phys. Rev. B {\bf 72}, 024544 (2005).
\bibitem{bayandin} K. V. Bayandin, G. B. Lesovik, and T. Martin,
Phys. Rev. B {\bf 74}, 085326 (2006). 
\bibitem{GHZ} D. M. Greenberger, M. A. Horne, A. Shimony and A. Zeilinger, American Journal of Physics \textbf{58}, 1131 (1990).
\bibitem{Cirel'son} B. S. Cirel'son, Lett. Math. Phys. \textbf{4}, 93 (1980).
\bibitem{NChuang} M. A. Nielsen and I. L. Chuang,  \textit{Quantum Computation and Quantum information} (Cambridge University Press, 2000).
\bibitem{tripartite rules 1} A. Karlsson and M. Bourennane, Phys. Rev. A \textbf{58}, 4394 (1998).
\bibitem{tripartite rules 2} J. C. Hao, C. F. Li and G. C. Guo, Phys. Rev. A \textbf{63},  054301 (2001).
\bibitem{zeilinger_ghz} A. Zeilinger, M.A. Horne, H. Weinfurter, and M. Zukowski, Phys. Rev. Lett. {\bf 78}, 3031 (1997). 
\bibitem{LebedevBlatter} A. V. Lebedev, G. Blatter, C. W. J. Beenakker and G. B. Lesovik, Phys. Rev. B  {\bf 69}, 235312 (2004)
\bibitem{koreans} Kicheon Kang and Kahng Ho Lee, Physica E {\bf 40}, 1395 (2008).
\bibitem{we} A. A. Vyshnevyy, A. V. Lebedev, G. B. Lesovik and G. Blatter (unpublished).
\bibitem{another 2 MZ} J. Dressel, Y. Choi, and A. N. Jordan, Phys. Rev. B \textbf{85}, 045320 (2012). 
\bibitem{3_photons} Dik Bouwmeester, Jian-Wei Pan, Matthew Daniell, Harald Weinfurter and Anton Zeilinger, 
Phys. Rev. Lett. {\bf 82}, 1345 (1999).
\bibitem{6_photons} Chao-Yang Lu, Xiao-Qi Zhou, Otfried G\"uhne, Wei-Bo Gao, Jin Zhang, Zhen-Sheng Yuan, Alexander Goebel, 
Tao Yang and Jian-Wei Pan, Nature Physics {\bf 3}, 91 (2007).   
\bibitem{shor_nmr} Lieven M. K. Vandersypen, Matthias Steffen, Gregory Breyta, Costantino S. Yannoni, Mark H. Sherwood and Isaac L. Chuang,
Nature {\bf 414}, 883 (2001).
\bibitem{glattli_single} G. F\`eve {\it et al.}, Science {\bf 316}, 1169 (2007). 
\bibitem{mach_zehnder}
Yang Ji, Yunchul Chung, D. Sprinzak, M. Heiblum, D. Mahalu and Hadas Shtrikman
Nature {\bf 422}, 415 (2003); 
I. Neder, N. Ofek, Y. Chung, M. Heiblum, D. Mahalu and V. Umansky, Nature {\bf 448}, 333 (2007);
P. Roulleau, F. Portier, D. C. Glattli, P. Roche, A. Cavanna, 
G. Faini, U. Gennser and D. Mailly, Phys. Rev. B {\bf 76}, 161309 (2007).
\bibitem{mahe} A. Mah\'e, F. D. Parmentier, E. Bocquillon, J.-M. Berroir, D. C. Glattli, T. Kontos, B. Placais, G. F\`eve, A. Cavanna and Y. Jin,
Phys. Rev. B {\bf 82}, 201309(R) (2010).   
\bibitem{Lesuslatter} G. B. Lesovik, M. V. Suslov and G. Blatter, Phys. Rev. A {\bf 82}, 012316 (2010) 
\bibitem{UFN_Klyshko}A. V. Belinskii and D. N. Klyshko, Phys. Usp. \textbf{36}, 653-693~(1993)
\bibitem{jonckheere} M. Albert, C. Flindt and M. B\"uttiker,
Phys. Rev. B {\bf 82}, 041407(R) (2010); 
M. Albert, C. Flindt and M. B\"uttiker,
Phys. Rev. Lett. {\bf 107}, 086805 (2011); G. Haack, M. Moskalets, J. Splettstoesser and M. B\"uttiker,
Phys. Rev. B {\bf 84}, 081303(R) (2011); T. Jonckheere, T. Stoll, J. Rech and T. Martin, Phys. Rev. B {\bf 85}, 045321 (2012).
\bibitem{MZ_dephasing} H. Forster, S. Pilgram and M. B\"uttiker, 
Phys. Rev. B {\bf 72}, 075301 (2005); 
F. Marquardt, Europhys. Lett. {\bf 72}, 788 (2005); 
F. Marquardt and C. Bruder, Phys. Rev. Lett. {\bf 92}, 056805 (2004). 
S.-C. Youn, H.-W. Lee and H.-S. Sim, Phys. Rev. Lett. {\bf 100}, 196807 (2008)
\bibitem{kovrizhin} D. L. Kovrizhin and J. T. Chalker,
Phys. Rev. B {\bf 81}, 155318 (2010).  
\bibitem{Glattli_dephasing} P. Roulleau, F. Portier, P. Roche, A. Cavanna, 
G. Faini, U. Gennser and D. Mailly, Phys. Rev. Lett. {\bf 101}, 186803 (2008). 
%\bibitem{Lorentz_on_demand}L. S. Levitov, H. Lee and G. B. Lesovik, 
%Journ. Math. Phys., \textbf{37}, issue 10, 4845-4866 (1996).
\bibitem{beenakker2} C.W.J. Beenakker, C. Emary and M. Kindermann,
Phys. Rev. B {\bf 69}, 115320 (2004)
\bibitem{reznik} G. Gershon, Yu. Bomze, E. V. Sukhorukov and M. Reznikov, Phys. Rev. Lett. {\bf 101}, 016803 (2008)
\bibitem{baykos} K. V. Bayandin, A.V. Lebedev and G. B. Lesovik,  JETP {\bf 106}, 117 (2008)
\end{thebibliography}
\end{document}